\documentclass[journal,comsoc]{IEEEtran}
\usepackage[T1]{fontenc}

\ifCLASSINFOpdf
\else
\fi
\usepackage{amsmath}
\usepackage[cmintegrals]{newtxmath}
\ifodd 1
\else

\fi

\usepackage{algpseudocode}

\usepackage{graphicx}
\usepackage{epstopdf}
\usepackage{subfigure}
\usepackage{amsmath}
\usepackage{bm}
\usepackage{color}
\usepackage{algorithm}
\usepackage{algpseudocode}

\usepackage{multirow}

\begin{document}
%
\title{Deep Reinforcement Learning with Spatio-temporal Traffic Forecasting for Data-Driven Base Station Sleep Control}

\author{
	Qiong Wu, Xu Chen,~\IEEEmembership{Senior Member,~IEEE}, Zhi Zhou,~\IEEEmembership{Member,~IEEE}, Liang Chen, Junshan Zhang,~\IEEEmembership{Fellow,~IEEE}
	\thanks{Q. Wu, X. Chen and Z. Zhou are with School of Computer Science and Engineering, Sun Yat-sen University, Guangzhou 510006, China. L. Chen is with the Department of Financial Technology, Tencent, Shenzhen 518054, China. J. Zhang is with the School of Electrical, Computer and Energy Engineering, Arizona State University, Tempe, AZ 85287 USA.}}

\maketitle

\begin{abstract}
To meet the ever increasing mobile traffic demand in 5G era, base stations (BSs) have been densely deployed in radio access networks (RANs) to increase the network coverage and capacity. However, as the high density of BSs is designed to accommodate peak traffic, it would consume an unnecessarily large amount of energy if BSs are on during off-peak time. To save the energy consumption of cellular networks, an effective way is to deactivate some idle base stations that do not serve any traffic demand. In this paper, we develop a traffic-aware dynamic BS sleep control framework, named DeepBSC, which presents a novel data-driven learning approach to determine the BS active/sleep modes while meeting lower energy consumption and satisfactory Quality of Service (QoS) requirements. Specifically, the traffic demands are predicted by the proposed GS-STN model, which leverages the geographical and semantic spatial-temporal correlations of mobile traffic. With accurate mobile traffic forecasting, the BS sleep control problem is cast as a Markov Decision Process that is solved by Actor-Critic reinforcement learning methods. To reduce the variance of cost estimation in the dynamic environment, we propose a benchmark transformation method that provides robust performance indicator for policy update. To expedite the training process, we adopt a Deep Deterministic Policy Gradient (DDPG) approach, together with an explorer network, which can strengthen the exploration further. Extensive experiments with a real-world dataset corroborate that our proposed framework significantly outperforms the existing methods.
\end{abstract}


%
\IEEEpeerreviewmaketitle

\section{Introduction}
The past decade has witnessed an explosive growth of mobile data traffic, which has triggered an accelerating deployment of base stations (BSs) to improve the cellular system capacity and enhance the network coverage. The deployment of numerous BSs incurs dramatic energy consumption. It has been reported that Information and Communication Technology (ICT) is becoming a significant part of the world energy consumption and expected to grow even further in the future \cite{marsan2009optimal}. For example, Telecom Italia, the incumbent telecommunications operator in Italy, consumes about $1\%$ of the total national energy demand, second only to the Italian railway system. In short, cellular networks are among the main energy consumers in the ICT field and BSs are responsible for over $80\%$ of the cellular network energy consumption \cite{richter2009energy}. Therefore, both researchers and operators are paying much attention to improving energy efficiency of BSs for sustainable development.

In cellular networks, the traffic load could have both temporal and spatial fluctuations which are dependent on time, location, and population distributions. Besides, in dense urban areas, base stations are deployed close to each other and there exist coverage overlappings among these BSs. Based on the facts above, dynamically switching the operation modes of BSs between active and sleep modes based on the traffic load fluctuation is a promising way to reduce energy consumption \cite{ZhengCCLZ15}. However, when some of the BSs are switched off to save energy, the user-perceived delay may then deteriorate due to traffic redirection. Therefore, one should carefully balance the tradeoff between energy saving and degradation in Quality of Service (QoS) when making BS active/sleep decisions \cite{8876870, 8463562}. Moreover, BS sleep control suffers from several practical constraints such as setup time and BS mode-changing cost (e.g., service migration cost, service delay, hardware wear-and-tear), indicating that the BS mode operation should be done in a time scale much slower than user association in practice.

Existing works on BS sleep operation rely heavily on idealistic assumptions such as Possion traffic demand models \cite{WuZN13} and threshold-based sleep control \cite{Kamitsos2010Optimal, Heyman1968}, which prevent it from being applied to complex realistic environments. Observing that real-world mobile traffic demands exhibit remarkable spatial-temporal correlations, many existing studies propose to utilize machine learning based methods (e.g., ARIMA \cite{KimLCCL11}) and deep learning approaches (e.g., STN \cite{ZhangP18}) for mobile traffic forecasting. Nevertheless, these approaches only focus on the temporal and geospatial correlations of mobile traffic demands without considering the fact that two areas with similar traffic patterns are not necessary to be geographically close. Thus, in this paper we construct a semantic spatial correlation graph and devise an efficient geographical and semantic spatial-temporal network (GS-STN) for timely and exact mobile traffic forecasting.

As for dynamic BS sleep control, it manifests a sequential decision-making process in nature \cite{LiuKZN18}. Any two consecutive BS switching operations are correlated with each other and the current BS switching operation will also further influence the overall energy consumption in the long run. However, many proposed analysis based approaches \cite{WuZN13} and greedy algorithms \cite{SonKYK11} ignore the sequential dependencies among the consecutive BS sleep control decisions. Moreover, they usually need repetitive and intensive computation at each time slot and thus are not suitable for practical BS sleep control in large-scale cellular networks. Recently, reinforcement learning (RL) has been proved to be a powerful tool to address this kind of complex sequential decision-making problems \cite{SuttonB98}. Nevertheless, we caution that traditional tabular-based RL methods would fail in large-scale systems as it would face the curse of dimensionality and the lack of exploration due to the large state-action space.

To tackle this challenge, we formulate BS sleep control as a Markov Decision Process (MDP) and then combine the recent advances in deep learning and Actor-Critic (AC) architecture to solve the MDP. Specifically, we take advantage of Deep Deterministic Policy Gradient (DDPG) algorithm in the training process. Note that the dynamic nature of the underlying network brings direct challenge to precise cost estimation in RL. For example, as the mobile traffic demand fluctuates dramatically with time, the cost estimation in RL can have a large variance. When the agent receives a decrease of cost, it is hard to distinguish if the decrease is the consequence of previous actions or environment changes. To address this issue, it is desirable to use the gap of the expected cost between the current policy and the optimal policy when evaluating the learning performance. Since the optimal policy is hard to obtain, we propose the benchmark transformation method which takes the cost estimated by some baseline policies (e.g., greedy algorithm) as benchmark and uses the cost gap as the criterion indicator. Furthermore, to assist exploration, we propose to add an explorer network, which can overcome the inefficiency of the conventional randomness-based exploration methods and significantly enhance the learning performance.

We summarize the main contributions of this paper as follows:
\begin{itemize}		
	\item We advocate a novel DeepBSC framework, which conducts traffic-aware dynamic sleep control, based on deep learning. The DeepBSC framework is data-driven and model-free, and is able to be applied to large-scale cellular networks with complex system dynamics.
	
	\item We devise the GS-STN model which harnesses the power of Convolutional Neural Network (CNN) and Long Short-Term Memory (LSTM) in a joint model, to capture the complex spatial-temporal correlations of mobile traffic demands. It should be noted that, for spatial correlations, we consider not only the correlations in geographical space, but also the correlations in semantic space by constructing a traffic similarity graph.
	
	\item We propose a deep reinforcement learning (DRL) approach to solve the dynamic BS sleep control problem. To reduce the variance of cost estimation caused by the varying traffic demands, we propose a novel benchmark transformation strategy.  Moreover, we also devise an explorer network to strengthen the exploration during the model training process further. Through these enhancements, we can achieve significant performance improvement and outstanding learning acceleration.
	
	\item Extensive experiments are conducted using a realistic dataset of Milan city, which demonstrates the efficiency of our proposed DeepBSC framework. Specifically, with our precise mobile traffic demands predicted by GS-STN, we can obtain $20.5\%$ cost reduction comparing with the widely-used ARIMA traffic forecasting model. Moreover, our DRL based BS sleep control method can achieve $12.8\%$ cost saving when adopting benchmark transformation and explorer network than the popular DQN method, which shows the effectiveness of our proposed DeepBSC framework.
\end{itemize}

The rest of this paper is organized as follows: In Section \ref{RelatedWork}, we review the related work of mobile traffic forecasting and dynamic BS sleep control. In Section \ref{SecFramework}, we introduce the system model and describe the problem formulation. And we also give the framework overview. The details of the GS-STN approach to predict mobile traffic demands are described in Section \ref{SecForecasting}. Our traffic-aware dynamic BS sleep control model is elaborated in Section \ref{BSsleeping}. We conduct extensive experiments in Section \ref{SecExperiments} and conclude our paper in Section \ref{SecConclusion}.

\section{Related Work}
\label{RelatedWork}
\subsection{Mobile Traffic Forecasting}
The recent works \cite{OhK10} and \cite{zhou2009green} present dynamic BS switching algorithms with the traffic loads as a prior and preliminarily demonstrate the effectiveness of energy saving. Wu et al. assume that the traffic demands follow Poisson process and then provide systematic insights  \cite{WuZN13}. However, these idealistic assumptions make these works suffer in practical applications as the real-world traffic loads are often much more complex due to the phenomenons such as self-similarity and non-stationarity \cite{LiuKZN18}. To dynamically and timely adjust the working status of BSs, many studies begin to conduct mobile traffic forecasting. For example, ARIMA is employed to predict data traffic across 9,000 base stations in Shanghai \cite{KimLCCL11}. Tikunov and Nishumura introduce a Holt-Winters exponential smoothing scheme for mobile traffic forecasting \cite{Tikunov2007Traffic}. Nevertheless, these existing mobile traffic prediction mechanisms only consider prior temporal information in individual region, while ignoring the important spatial correlations of the traffic demands in adjacent regions. Accordingly, spatio-temporal patterns of mobile traffic have been recently considered \cite{HuangCL17}. However, these methods are only limited to geospatial correlations based on physical proximity. Different from these works, we propose to leverage the spatial correlations in both dimensions of geography and semantics, which demonstrates excellent performance in our mobile traffic forecasting problem.

\subsection{Base Station Sleep Control}
The problem of energy saving with dynamic BS switching is a well-known combinatorial problem, which has been proven to be NP-hard \cite{WongYP12}. Solving such problem generally requires global information (e.g., mobile traffic demand information), which makes it more challenging. To address this problem, some greedy and heuristic algorithms are proposed. For example, Son et al. develop greedy-on and greedy-off algorithms for BS energy saving \cite{SonKYK11}. Furthermore, there are also some methods that tackle the BS operation problem with optimization approaches \cite{LiaoHLL14, ZhuangGH16}. These methods can find the optimal or sub-optimal configuration of BS modes based on the assumption that the network environment remains unchanged in the considered period. For the simplistic setting of Poisson traffic demand and exponential service time, the double-threshold hysteretic policy has been proven to be optimal \cite{Kamitsos2010Optimal, Heyman1968}. While for more general traffic and service patterns, the double-threshold policy is not guaranteed to be optimal. Moreover, it is computationally-challenging to determine the proper threshold values \cite{leng2017wait}. Li et al. formulate the traffic variations as a Markov decision process and design a reinforcement learning framework based BS mode switching scheme \cite{LiZCPZ14}. Liu et al. consider a single BS sleep control problem with varying traffic patterns using Deep Q-Network (DQN) \cite{LiuKZN18}. Nevertheless, these approaches can only handle low-dimensional action spaces and are likely intractable for high-dimensional action spaces which are difficult to explore efficiently. In this paper, we hence propose a traffic-aware DRL-based BS sleep control approach for large-scale networks. We introduce the ideas of benchmark transformation and explorer network to significantly enhance the learning performance.

\section{Traffic-Aware Sleep Control for Base Stations}
\label{SecFramework}
\subsection{Network Model}
\begin{figure}[!t]
	\centering
	\includegraphics[width=0.9\linewidth]{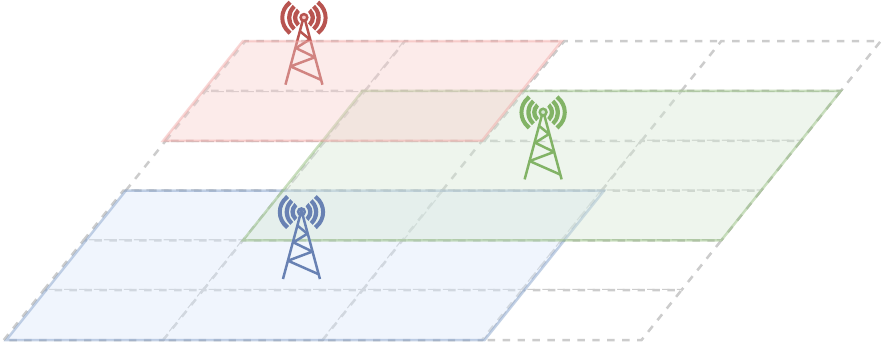}
	\caption{Network scenario. Each base station is deployed to handle the traffic demand in its coverage region and different base stations may have coverage overlaps.}
	\label{example}
	\vspace{-15pt}
\end{figure}

A cellular network usually consists of multiple base stations (BSs) to handle the mobile traffic loads. In this paper, we consider a cellular network served by a set of BSs, denoted $\mathcal{B} = \{1, 2, \cdots, B\}$. As depicted in Fig. \ref{example}, the geographical region can be divided into non-overlapping grids, each base station is deployed to handle traffic loads in its coverage region (e.g., $3 \times 3$ grids) and different base stations may have coverage overlaps. As mobile traffic fluctuates over time, many base stations are under-utilized most time, which would result in significant energy wastage and heavy energy inefficiency. Thus, it is necessary to change the operation modes of BSs dynamically according to the varying traffic demands. We assume that each base station is equipped with a traffic monitor that can sense the traffic load in its coverage area and correspondingly determine the operation mode (e.g., active/sleep mode) of the base station according to the traffic volume. However, frequent changes of the BS modes will bring high cost of service migration, service delay and hardware wear-and-tear, thus we tend to perform BS active/sleep mode operation in a large time scale. Specifically, we discretize the time into time slots, and each time slot has a span of half an hour during which the mode of base station remains unchanged.

\subsection{Base Station Cost Model}
\textbf{Energy Consumption:} In cellular network, the energy consumption of a base station is not simply proportional to the traffic loads within its coverage \cite{LiZWZZ12}. In fact, the energy consumption of BS consists of two categories: fixed energy consumption that is irrelevant to BS's traffic loads and load-related energy cost. The fixed cost comes from circuit consumption and cooling consumption, while the load-related energy cost comes from the power amplifier. Hence, we adopt the generalized energy model to represent the cost of a BS $i$ at time slot $t$, which can be summarized as
\begin{equation}
p_{i}^{t} = P_{i}^{f} + P_{i}^{l}(\rho_{i}^{t} ),
\end{equation}
where $P_{i}^{f}$ is the fixed energy consumption and $P_{i}^{l}$ is load-dependent energy cost which is relevant to $\rho_{i}^{t}$, the traffic load of BS $i$ at time slot $t$.

\textbf{QoS degradation cost:} Deactivating BSs will cause the degradation of user's Quality of Service (QoS). Specifically, user may suffer longer transmission delay and service delay. The transmission delay can be denoted as
\begin{equation}
c_{tran}^{t}(D^{t},a^{t}) = f_{tran}(D^{t}, a^{t}),
\end{equation}
where $D^{t}$ is the mobile traffic volume in all regions and $a^{t}$ indicates the active/sleep modes of base stations at time slot $t$. $f_{tran}(\cdot)$ is the traffic dispatching function defined by the network operator to reallocate the mobile traffic demands in different regions to the active base stations. Here our model allows the flexibility that different network operators can implement different traffic management schemes tailored to their own operation demands and requirements (e.g., the trade-off between QoS and energy consumption). In our experiment, we use the widely adopted minimum cost network flow based traffic routing \cite{KangSZ12} as our traffic dispatching scheme. The total service delay of all base stations can be represented as:
\begin{equation}
c_{ser}^{t}(\rho^{t},C^{t}) = \sum_{i \in \mathcal{B}} f_{ser}(\rho_{i}^{t}, C_{i}^{t}),
\end{equation}
where the service latency of a base station $i$ is a function that depends on its traffic load $\rho_{i}^{t}$ and its current service capacity $C_{i}^{t}$. For ease of implementation, we use the average queueing delay \cite{newell2013applications} to determine the function as $f_{ser}(\rho_{i}^{t}, C_{i}^{t}) =  \frac{1}{C_{i}^{t}-\rho_{i}^{t}}$.
And thus the QoS degradation cost can be summarized as
\begin{equation}
c_{d}^{t}(D^{t},a^{t},\rho^{t},C^{t}) = \beta_{d} \cdot (c_{tran}^{t} (D^{t},a^{t}) + c_{ser}^{t}(\rho^{t},C^{t}) ),
\end{equation}
where $\beta_{d}$ is a penalty factor.

\begin{table}[!t]
	\small
	\vspace{5pt}
	\caption{Key notation. }
	\label{notations}       
	\newcommand{\tabincell}[2]{\begin{tabular}{@{}#1@{}}#2\end{tabular}}
	\centering
	\begin{tabular}{|c|c|}
		\hline Notations & Definitions  \\
		\hline	\hline $D^{t}$ & Mobile traffic volume at time slot t \\
		\hline $\mathcal{B}$ & The set of base stations (BSs)\\
		\hline $B$ & The number of base stations \\
		\hline $p_{i}^{t}$ & Energy consumption of BS i at time slot $t$\\
		\hline $\rho_{i}^{t}$ & Traffic load of BS i at time slot $t$\\
		\hline $\beta_{d}$ & Penalty factor of QoS degradation\\
		\hline $\beta_{s}$ & Penalty factor of switching cost\\
		\hline $a^{t}$ & Active/sleep modes of BSs at time slot $t$\\
		\hline $s^{t}$ & \tabincell{c}{The state (including mobile traffic and   \\active/sleep modes) in time slot $t$} \\
		\hline $c(s^{t},a^{t})$ & Cost estimated by our agent given $s^{t}$ and $a^{t}$\\
		\hline $g(s^{t}, a^{t})$ & \tabincell{c}{Cost gap between $c(s^{t},a^{t})$ and \\benchmark cost}\\
		\hline $\pi(s^{t}|\theta_{\pi})$& Actor network with parameters $\theta_{\pi}$ \\
		\hline $Q(s^{t}, a^{t}|\theta_{Q})$ & Critic network with parameters $\theta_{Q}$ \\
		\hline
	\end{tabular}
\end{table}

\textbf{Base station switching cost:} The base station switching cost is incurred by toggling base stations into or out of a power-saving mode between two adjacent time slots and includes the service migration, service delay and hardware wear-and-tear costs. Let $\beta_{s}^{i}$ be the cost to toggle the base station $i$ from sleep mode to the active mode and we assume the cost of transiting from the active to the sleep mode is $0$. If this is not the case, we can simply fold the corresponding cost into the cost $\beta_{s}^{i}$ incurred in the next power-up operation. Then the switching cost for changing the operation modes of base stations from $a^{t-1}$ to $a^{t}$ is
\begin{equation}
c_{s}^{t}(a^{t-1}, a^{t}) =\sum_{i \in \mathcal{B}} \beta_{s}^{i} \cdot (a_{i}^{t}-a_{i}^{t-1})^{+},
\end{equation}
where the function $(x)^{+}$ is defined as
$$
(x)^{+} = \left\{
\begin{aligned}
&x,& & {if \ x > 0},\\
&0,& & {else}.\\
\end{aligned}
\right.
$$
To make it clear, the key notations used in this paper are summarized in Table \ref{notations}.

\subsection{Problem Formulation}
The primary objective of this study is to find the optimal BS sleep control policy that can minimize the total system cost in the long run. We formulate the dynamic BS active/sleep operation optimization problem as a Markov Decision Process (MDP), which can be described as a tuple $\mathcal{M} = <\mathcal{S}, \mathcal{A}, \mathcal{P}, \mathcal{C}, \gamma>$ wherein $\mathcal{S}$ is the state space, $\mathcal{A}$ is the action space, $\mathcal{P}$ is the state transition probability function, $\mathcal{C}$ is the cost function and $\gamma \in [0, 1]$ is the discount factor. We introduce the MDP design in detail as follows.
\subsubsection{State} At each time slot $t$, the system state is composed of the mobile traffic volume $D^{t}$ and the BS active/sleep modes $a^{t-1}$ in previous time slot $t-1$.  However, the traffic volume $D^{t}$ is unavailable at the beginning of time slot $t$ when the BS active/sleep mode operation is to be made. Alternatively, we use the predicted value $\tilde{D}^{t}$, which can be obtained by the traffic prediction model we proposed. Thus, the state can be represented as  $s^{t} = (\tilde{D}^{t}, a^{t-1})$.
\subsubsection{Action and Transition Function}The action $a^{t}$ is the decision of active/sleep modes for all the base stations at time slot $t$. Given the current state $s^{t}$ and action $a^{t}$, the distribution of the next state $s^{t+1}$ can be represented by the state transition probability function $P(s^{t+1}|s^{t},a^{t})$.

\subsubsection{Cost Function} The overall cost of the system at time slot $t$ is composed of energy consumption of BSs, user's QoS degradation cost and BS switching cost:
\begin{equation}
c^{t} = \sum_{i \in \mathcal{B}}p_{i}^{t} + c_{d}^{t}(D^{t},a^{t}, \rho^{t},C^{t}) + c_{s}^{t}(a^{t-1},a^{t}).
\end{equation}

At each time slot, we aim to find the action (BS active/sleep modes) which can minimize the long-term system cost. To achieve this goal, a major branch of reinforcement learning approaches focus on estimating the value function, among them Q-learning has attracted lots of attention for its simplicity and effectiveness \cite{DBLP:journals/ras/Krose95}. Q-learning uses the state-action value function $Q(s^{t},a^{t})$, which is defined as the expected value of accumulated cost starting from the current state when taking the action $a^{t}$, for the estimation of Q value. The state-action value function is given by
\begin{equation}
Q(s^{t},a^{t}) = c^{t} + \gamma \cdot min_{a^{\prime}}\mathbb{E}[Q(s^{t+1},a^{\prime})].
\end{equation}
It contains two parts: the immediate cost generated in current time slot and the discount state-action value function of a subsequent state. This recursive relationship is called the Bellman equation, which can be used to calculate the true value of $Q(s^{t},a^{t})$. Then the problem becomes:
\begin{equation}
\begin{split}
& \min_{a} Q(s^{t},a), \\
&s.t. \quad a_{i} \in \{0,1\},\, \forall i, s^{t}.
\end{split}
\end{equation}
Our objective is to find the action that solves the problem in Eqn. (8) given each system state $s^{t}$.

\begin{figure}[!t]
	\centering
	\includegraphics[width=0.8\linewidth]{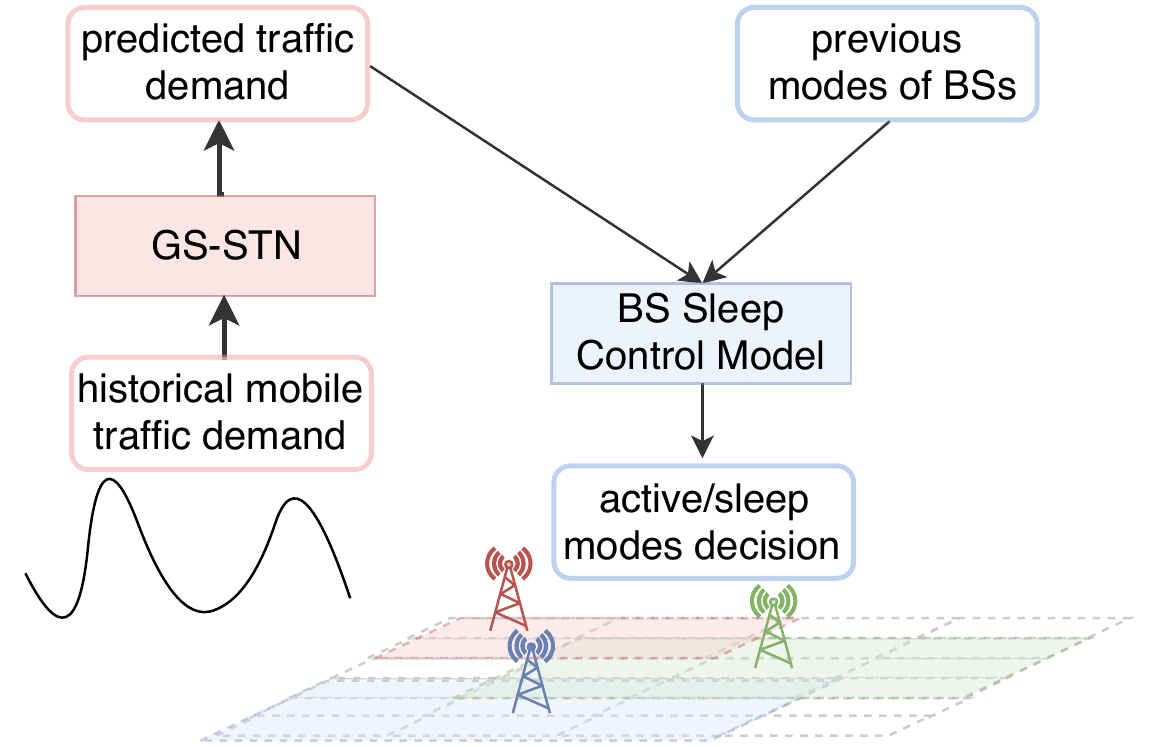}
	\caption{The framework overiew of our proposed DeepBSC model.}
	\label{overview}
	\vspace{-10pt}
\end{figure}

As mentioned above, the mobile traffic demand is unavailable at the beginning of a time slot, thus in order to make wise BS active/sleep mode operation, we first predict the upcoming mobile traffic demand, which is crucial in the following BS sleep control step. In this paper, we propose a novel DeepBSC framework for traffic-aware energy-saving BS sleep control as shown in Fig. \ref{overview}. We first conduct precise mobile traffic forecasting using GS-STN approach which leverages the geographical and semantic spatial-temporal correlations of mobile traffic demands (see Section \ref{SecForecasting}). Given the predicted traffic demands and the active/sleep modes of BSs in previous time slot, our BS sleep control model will generate new modes for BSs in order to minimize the long-term system cost and we will describe the details in Section \ref{BSsleeping}.

\section{Geographical and Semantic Spatial-Temporal Network for Mobile Traffic Forecasting}
\label{SecForecasting}
\subsection{Mobile Traffic Forecasting}
The geographical area of a city can be divided into $X \times Y$ non-overlapping grids, then the mobile traffic volume in this area at time slot $t$ is
\begin{equation}
\quad D^{t} = \begin{bmatrix}
d^{t}_{(1,1)} & \cdots\ & d^{t}_{(1,Y)}\\
\vdots & \vdots & \vdots\\
d^{t}_{(X,1)} & \cdots & d^{t}_{(X,Y)}\\
\end{bmatrix},
\end{equation}
where $d^{t}_{(x,y)}$ measures the data traffic volume in a grid with coordinates $(x,y)$. To make accurate mobile traffic forecasting, we aim to find the most likely traffic demand at time slot $t$ given truncated historical demand data
\begin{equation}
\tilde{D}_{t} = \arg\max_{D_{t}} \; p(D_{t}|D_{t-K}, ..., D_{t-1} ),
\end{equation}
where $D_{t-K}, ..., D_{t-1}$ are the observations of mobile traffic demand at the previous $K$ time slots. As the traffic patterns are complex non-linear in nature, we propose to use deep learning tools to model the marginal distribution above. That is, we aim to find the function $\mathcal{F}(\cdot)$ which can output the accurate mobile traffic demand
\begin{equation}
\tilde{D}_{t} = \mathcal{F}(D_{t-K}, ..., D_{t-1}).
\end{equation}

\subsection{Geographical and Semantic Spatial-Temporal Network}
\begin{figure}[!t]
	\centering
	\includegraphics[width=0.85\linewidth]{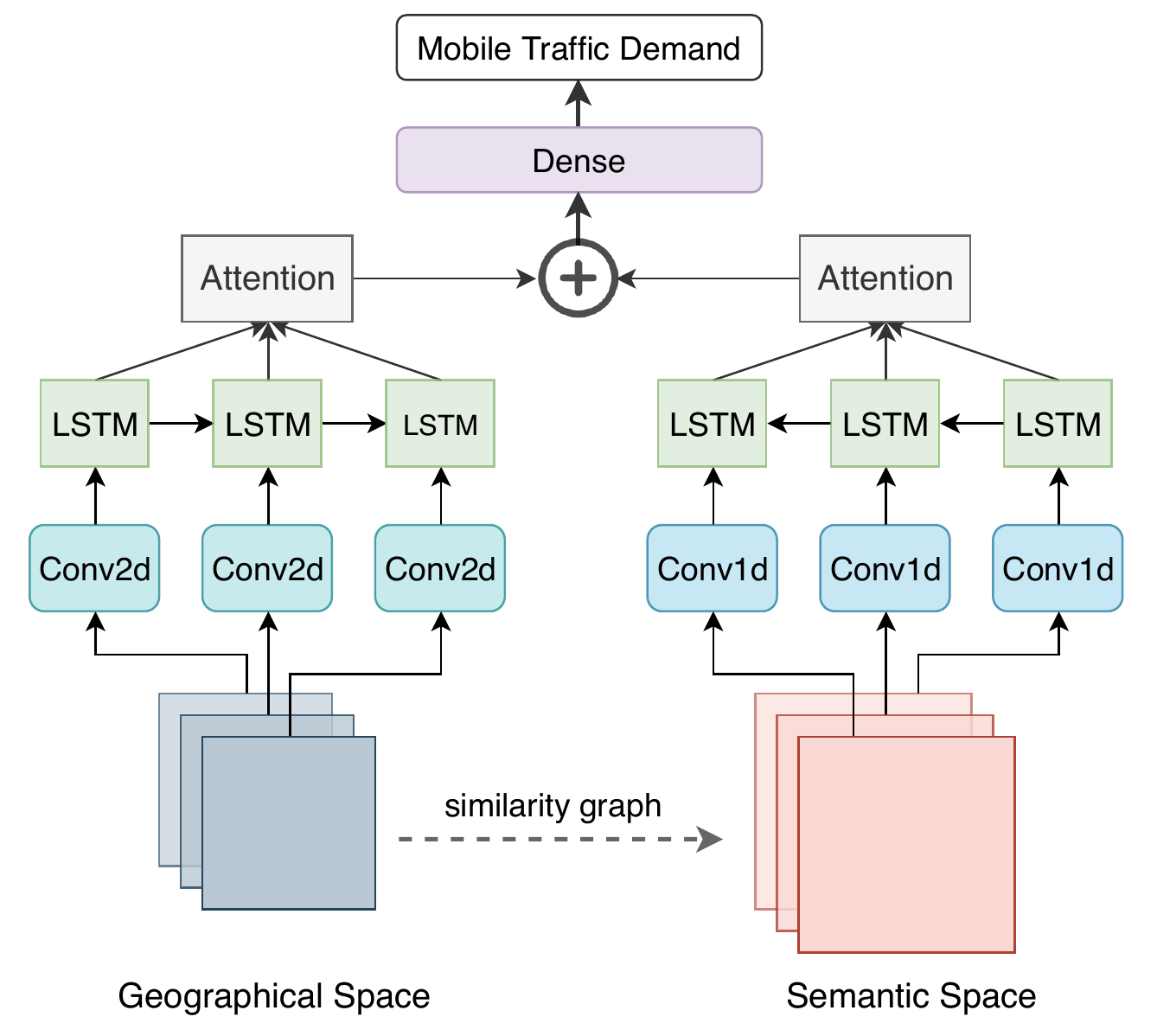}
	\caption{Geographical and semantic spatial-temporal network.}
	\label{STN}
	\vspace{-15pt}
\end{figure}

As shown in \cite{XuLWZJ17,FumoFS17}, traffic patterns exhibit strong spatio-temporal correlations. With this insight, we use CNN and LSTM to capture the complex relations of both space and time.  In particular, we consider spatial correlations in both dimensions of geography and semantics, and then propose Geographical and Semantic Spatial-Temporal Network (GS-STN) model as shown in Fig. \ref{STN}.

\textbf{Geospatial correlation:} Motivated by the First Law of Geography : ``near things are more related than distant things'',  many studies focus on mobile traffic forecasting using the information of neighboring regions while filtering the weakly correlated remote regions \cite{Yao0KTJLGYL18}. For geospatially nearby regions, we define that the mobile traffic demand $d^{t}_{(x,y)}$ is correlated with its surrounding $(r+1)\times(r+1)$ grid matrix $M$, where
\begin{equation}
\quad M = \begin{bmatrix}
d^{t}_{(x-\frac{r}{2},y-\frac{r}{2})} & \cdots\ & d^{t}_{(x-\frac{r}{2},y + \frac{r}{2})}\\
\vdots & d^{t}_{x,y} & \vdots\\
d^{t}_{(x+\frac{r}{2},y-\frac{r}{2})} & \cdots & d^{t}_{(x+\frac{r}{2},y+\frac{r}{2})}\\
\end{bmatrix},
\end{equation}
here, the parameter $r$ is always an even number and controls the spatial granularity. To capture the spatial correlations of nearby regions, we use CNN which is capable of modeling complex spatial interactions. Given the 2-D mobile traffic map $D^{t}$, we specify the operations of CNN in Eqn. (\ref{conv2d}),
\begin{equation}
\label{conv2d}
S^{t}_{Geo} = f(D^{t} * W_{Geo} + b_{Geo}),
\end{equation}
where $*$ denotes the 2-D convolution operation and f($\cdot$) is an activation function. In this paper, we use the rectifier function as the activation, i.e., $f(z) = max(0,z)$. $W_{Geo}$ and $b_{Geo}$ are learnable weights and bias in the convolution layer. Given the historical mobile traffic $\mathcal{D} = [D^{t-K}, ..., D^{t-1}]$, we can get $\mathcal{S}_{Geo} = [S_{Geo}^{t-K}, ..., S_{Geo}^{t-1}]$, which captures the geographical correlations of mobile traffic.

\textbf{Semantic spatial correlation:} Another key observation is that two areas with similar traffic patterns are not necessary to be geographically close. For example, rural areas in a city may exhibit similar traffic demand patterns while they are usually distant spatially as urban areas separate them. Inspired by this, we construct a semantic spatial correlation graph by calculating the cosine similarity among mobile traffic loads in all grids to capture the correlations of all the grids in semantic space. The traffic similarity graph can be denoted by a matrix $A\in \mathbb{R}^{N\times N}$  where N is the total number of grids. The  $i$-th row  describes the cosine similarity of mobile traffic loads between the $i$-th grid and all the N grids. We further apply 1-D convolution operation to aggregate all the traffic similarity information of other grids for each row. Given the historical mobile traffic information $\mathcal{D} = [D^{t-K}, ..., D^{t-1}]$, we can get the semantic correlation $\mathcal{S}_{Sem} = [S_{Sem}^{t-K}, ..., S_{Sem}^{t-1}]$.

\textbf{Temporal correlation:} To model sequential relations of mobile traffic demands, we propose to use Long Short-Term Memory (LSTM), which is a special recurrent neural network (RNN) that remedies the vanishing gradient problem characteristic to RNNs in long sequence training \cite{HochreiterS97, Informatik2001Gradient}. LSTM can learn sequential correlations stably by maintaining a memory cell $\bm{m}^{t}$ in time slot $t$, which can accumulate the previous sequential information. In addition, LSTM introduces three gates: $\bm{i}^{t}$, $\bm{o}^{t}$, $\bm{f}^{t}$, which decide the amount of new information to be added and the amount of historical information to be forgot. The details of LSTM can be founded in \cite{Informatik2001Gradient}. Furthermore, we enforce the LSTM model to focus on important parts through an attention mechanism as depicted in Fig. \ref{STN}. After conducting the attention operation, we can get the features $\mathcal{H}_{Geo}$ for $\mathcal{S}_{Geo}$ in geographical space and $\mathcal{H}_{Sem}$ for $\mathcal{S}_{Sem}$ in semantic space. We join all the features by concatenating them with the concatenation operator ``$\oplus$'' and then feed them to a fully-connected neural network to get the predicted mobile traffic demands in next time slot:
\begin{equation}
\tilde{D}^{t} = \sigma (W_{fc}(\mathcal{H}_{Geo} \oplus \mathcal{H}_{Sec}) + b_{fc}),
\end{equation}
where $W_{fc}$  and $b_{fc}$ are learnable parameters, and $\sigma(\cdot)$ is sigmoid function.

In conclusion, our proposed GS-STN model is composed of four subnetworks: (1) CNN network to capture geospatial correlations, (2) CNN network to capture spatial correlations in semantic space, (3) LSTM network to model temporal sequential relations of mobile traffic demands and (4) fully-connected neural network (dense layer) to concatenate all the features and predict the mobile traffic demands in the next time. Thus, the parameters of these subnetworks constitute the learnable parameters of our proposed GS-STN network. Moreover, we train these parameters in an end-to-end fashion. By considering the spatial-temporal correlations in both dimensions of geography and semantics, our proposed GS-STN approach can achieve significant performance improvement for mobile traffic forecasting comparing with other widely-used spatial-temporal learning methods (e.g., $8.4\%$ improvement than CNN-LSTM method in terms of NRMSE).

\section{Dynamic Base Station Sleep Control}
\label{BSsleeping}

\begin{figure}[!t]
	\centering
	\includegraphics[width=0.7\linewidth]{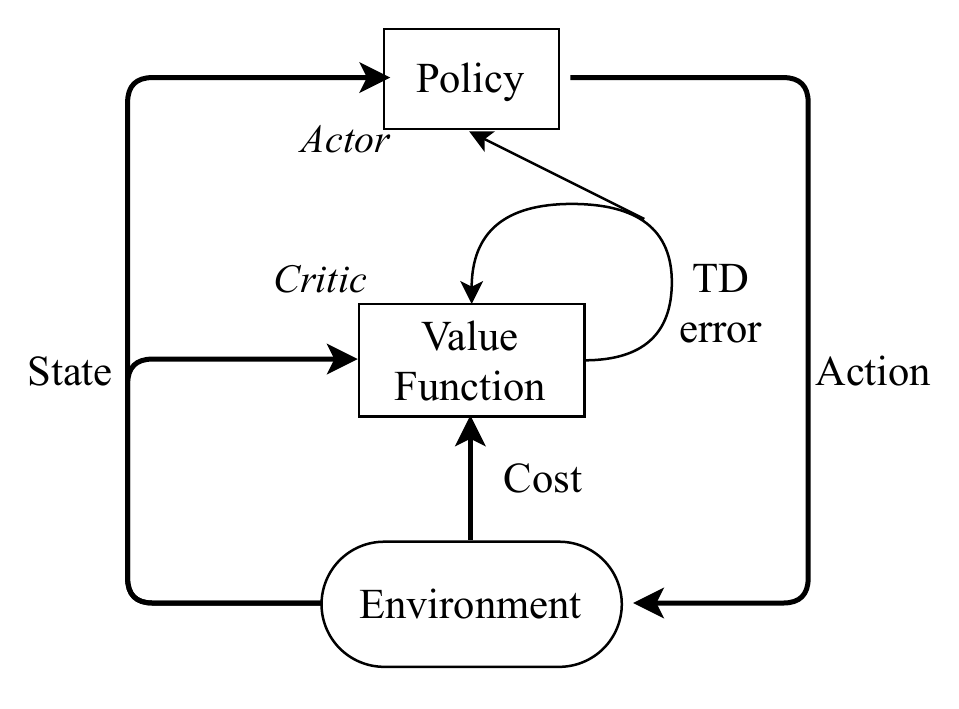}
	\caption{The actor-critic reinforcement learning framework.}
	\label{AC}
	\vspace{-15pt}
\end{figure}
\subsection{Actor Critic Architecture}
\label{ACarche}
There have been some well-known methods to solve the MDP problems such as dynamic programming (DP) \cite{SuttonB98}. Unfortunately, these DP approaches heavily rely on a well-defined mathematical model, e.g., the transition probabilities of the underlying system. However, it's challenging to know the intricate dynamics of the complex environment precisely. Therefore, in order to solve the dynamic base station sleep control problem, we employ reinforcement learning (RL) approaches that do not require any assumptions on the model and learn the model through interacting with the environment.

In this paper, we adopt the actor-critic (AC) framework, which inherits the advantages of both policy-based and value-based RL approaches, to solve the problem \cite{ye2018drag}. As illustrated in Fig. \ref{AC}, the actor-critic algorithm encompasses three components: actor, critic and the environment. At a given state, the actor, which is usually a parameterized policy, generates action and then executes it. This execution transforms the state of the environment into a new one and feeds back the corresponding cost to the critic. Then, the critic, which can be represented as a value function, criticises the action generated by the actor and refines the value function with the time difference (TD)-error. After the criticism, the actor will be guided by the TD-error to update the policy and then produce the action with a smaller cost. The algorithm repeats the above procedure until convergence.

As the actor-critic algorithm generates the action directly from the stored policy, it requires little computation to select an action to perform, which is critical in time-sensitive scenarios. However, classical AC approaches often suffer from the large storage space and inefficient learning in large-scale problems. The reason is that the critic often utilizes a state-action value table as the value function, which requires an exponentially growing storage with respect to the dimension of action and state variables. In experiment setting of our problem, there are one hundred grids and one hundred base stations. As a result, it's too large to learn the values for all state-action pairs separately and will hardly produce satisfactory performance. In addition, some state transitions may never happen (e.g., all the BSs go into sleep modes), which make it intractable for traditional tabular based RL approaches to learn the model parameters from certain state transitions as they lack the ability of prediction and generation.

To address the abovementioned issues, we utilize continuous parameterized functions to approximate the policy function and the value function \cite{WeiYSH18} instead of using tabular based approaches. Specifically, we adopt deep neural network (DNN) which has been identified as a universal function approximator that can approximate any function mapping, possibly with multiple inputs and outputs, as our actor and critic networks \cite{HornikSW89}.

The actor network directly outputs a single deterministic action vector to represent the base station mode operation. The actor network, also known as policy DNN, is given as
\begin{equation}
\tilde{a}^{t} = \pi(s^{t}|\theta_{\pi}),
\end{equation}
where $s^{t}$ is the state in time slot $t$ and $\theta_{\pi}$ is the parameters of the actor network. We adopt $sigmoid(x)$ as the activation of the output layer to confine the output values to $[0,1]$ which indicates the probability of the base stations to be activated or slept. $a^{t}$ is the final base station operation action obtained by mapping $\tilde{a}^{t}$ to 0-1 values.

The critic network, also known as value DNN, is to evaluate how good a policy is given a state-action pair. It can be represented as
\begin{equation}
q^{t} = Q (s^{t},a^{t}|\theta_{Q}),
\end{equation}
where the state $s^{t}$ and action $a^{t}$ are the inputs, $\theta_{Q}$ are the training parameters of the critic network. $q^{t}$ is the approximated Q-value of the state-action pair $(s^{t}, a^{t})$.

\subsection{Benchmark Transformation}
\begin{figure}[!t]
	\centering
	\includegraphics[width=0.9\linewidth]{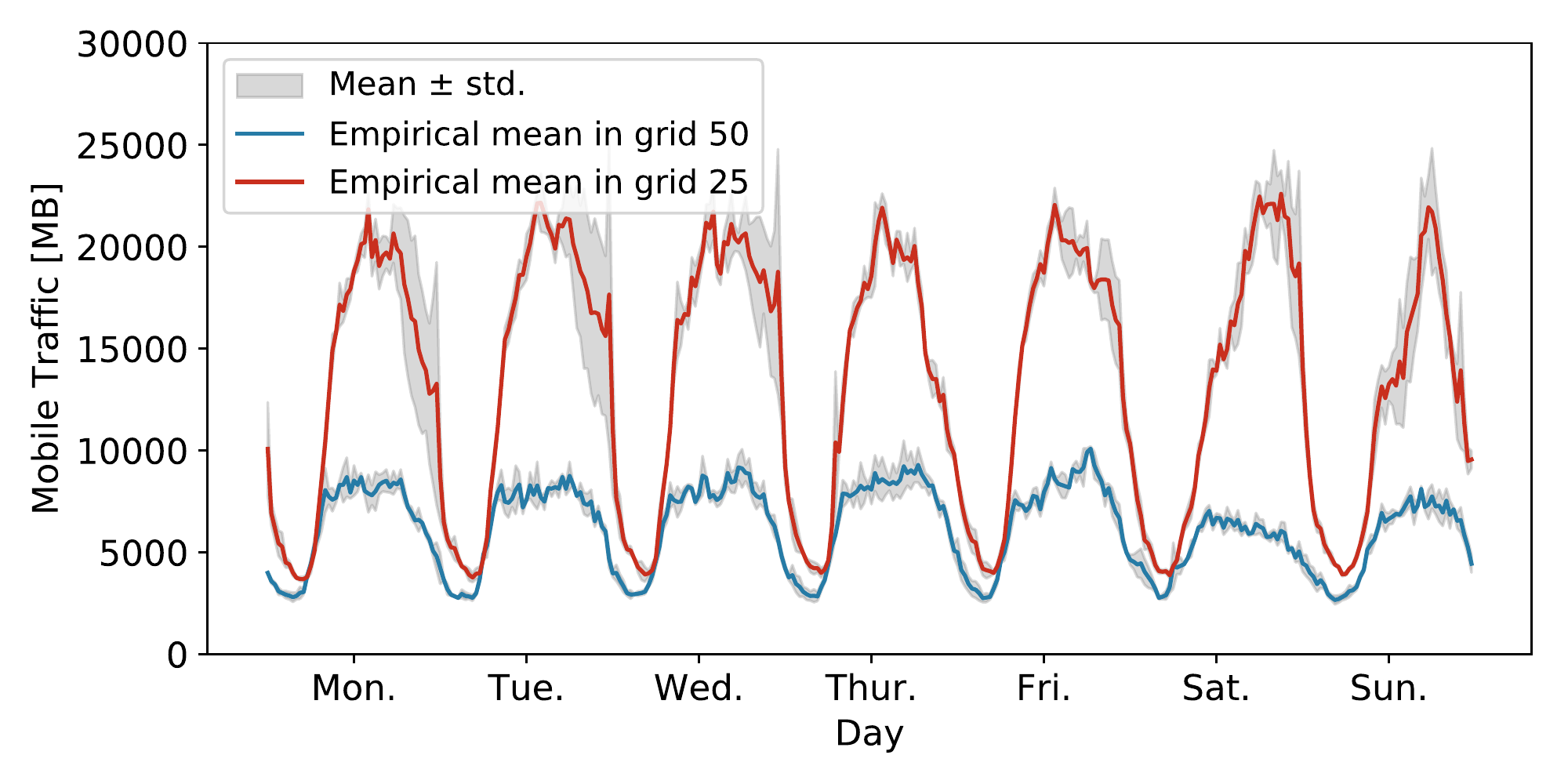}
	\caption{Mobile traffic demand sampled from $1^{st}$ Nov to $22^{th}$ Nov 2013. We calculate the empirical mean and standard deviation in grid 25 and grid 50, which shows the drastical fluctuation in both space and time.}
	\label{mobiletraffic}
	\vspace{-10pt}
\end{figure}

As shown in Fig. \ref{mobiletraffic}, the mobile traffic demand in a grid varies drastically at different time of a day, which means that there will be many changes and fluctuations in the dynamic environment of base station sleep control. Accordingly, it will cause a very large fluctuation in cost estimation which further affects the final learning performance. For example, the mobile traffic load in the traffic peak time is much higher than that in the off-peak time, which may cause the huge fluctuation of the total cost in the environment (i.e., energy cost, QoS degradation cost). However, when the agent receives a decrease of  the cost, it is unable to distinguish if the decrease is the consequence of previous actions or environment changes (e.g., lower mobile traffic demand).

To address this issue, instead of utilizing the absolute cost estimated by our agent which is susceptible to the dynamic environment changes, it is desirable to use the gap of the expected cost between current policy and optimal policy to evaluate the learning performance. Unfortunately, it's hard to obtain the optimal policy due to the large state-action space and dynamic environment. Instead, we propose a novel benchmark transformation strategy in which we rely on the expected difference between the cost estimated by current policy and the cost provided by baselines as the criterion indicator. Then, the goal is to learn how well our agent approaches or even exceeds the baselines. Specifically, at time slot $t$, we estimate the expected cost $ c^{t}_{base}(s^{t})$ using baseline methods (e.g., greedy algorithm) for the current state $s^{t}$ and the expected cost $c(s^{t},a^{t})$ under current policy of DeepBSC, thus the cost gap can be written as:
\begin{equation}
g(s^{t}, a^{t}) = c(s^{t},a^{t}) - c^{t}_{base}(s^{t}).
\end{equation}

\begin{figure}[!t]
	\centering
	\includegraphics[width=0.85\linewidth]{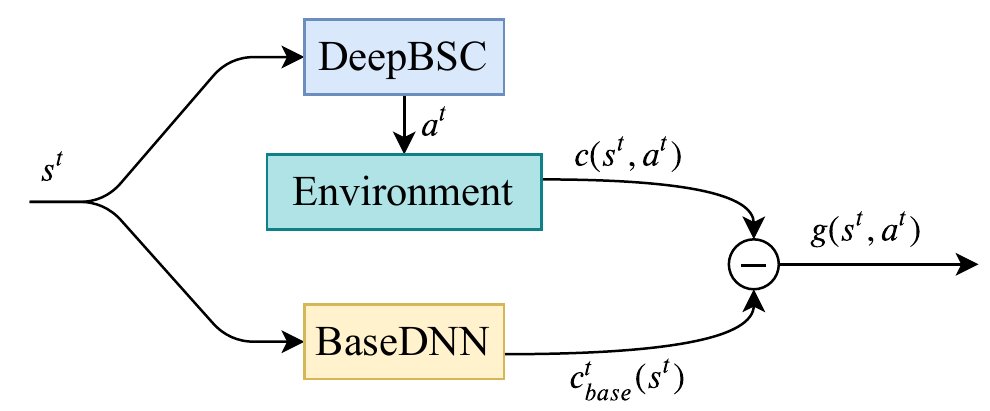}
	\caption{Details of benchmark transformation strategy.}
	\label{BR}
	\vspace{-10pt}
\end{figure}

In order to estimate the cost benchmark of different states timely and efficiently, we train an offline learning model, named BaseDNN, which takes the advantage of DNN with collected state-cost pairs $<s, c_{base}>$, and is able to output the cost benchmark for a given state online during the base station sleep control process. Note that we ignore the action taken by the baselines as the cost estimated by baselines can be seen as inherent cost of current environment. And the gap is the gain which caused by the action performed by our agent. The process of our proposed benchmark transformation strategy is depicted in Fig. \ref{BR}.

In the experiment, we choose the time scale of half an hour based on the need of BS sleep control, which is a suitable time granularity for control decision making in practice. When a smaller time scale is used, the magnitude of traffic demand fluctuations between two consecutive time slots could be smaller, but it still changes stochastically. Moreover, we need to predict the traffic demand for cost estimation and the prediction error exists inevitably. As a result, we still have the variance issue in cost estimation (although the impact of this issue can be reduced but not eliminated when using a smaller time scale). And hence the benchmark transformation is still useful for improving the learning performance. Nevertheless, if we use a very small time scale to run the algorithms, the implementation cost can be very significant since we need to conduct the traffic demand prediction and decision making at a very high frequency. In this regard, the benchmark transformation approach is more cost-effective.

\subsection{Explorer Network}
\begin{figure}[!t]
	\centering
	\includegraphics[width=0.95\linewidth]{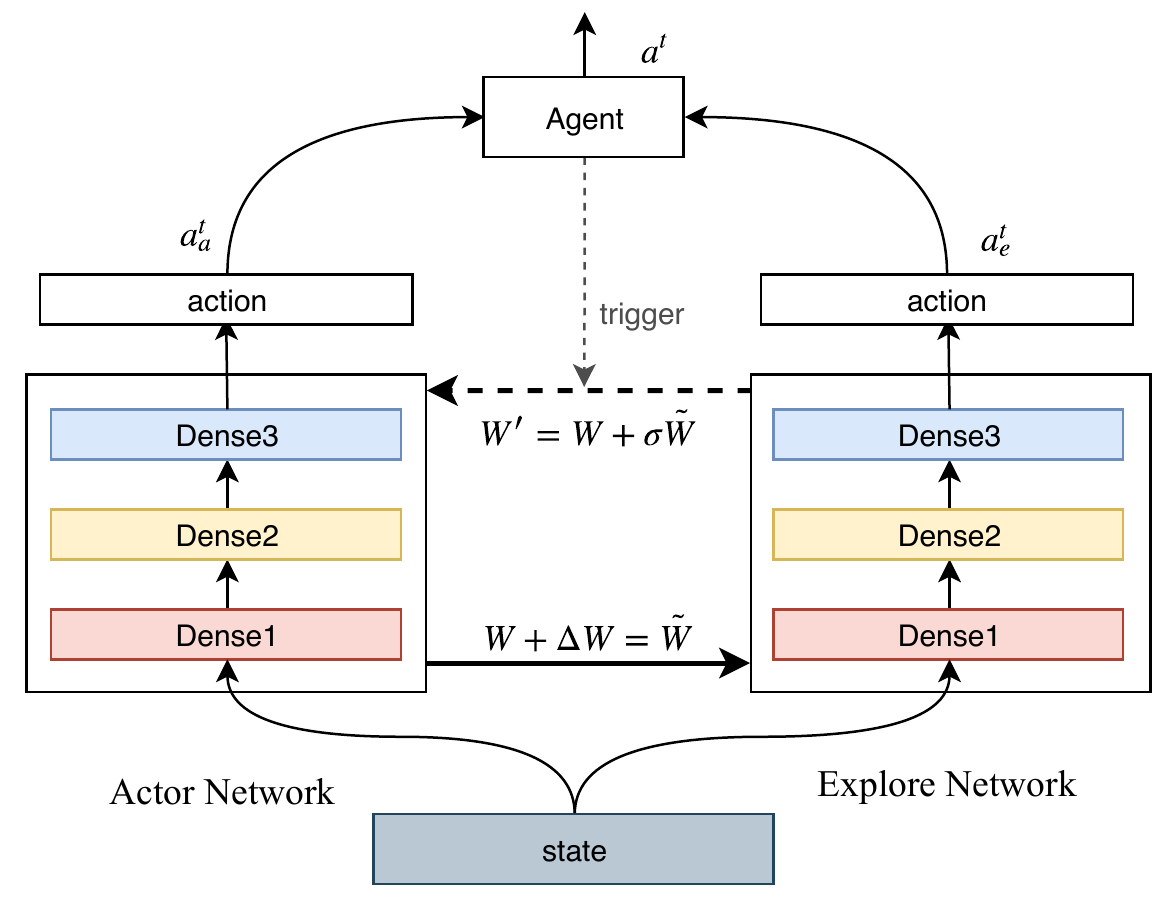}
	\caption{The explorer network. When the agent selects the action made by explorer network, then it will trigger the update of parameters in actor network towards explorer network.}
	\label{EN}
	\vspace{-10pt}
\end{figure}

As mentioned in Section \ref{ACarche}, our actor network deterministically outputs a single action to reduce complexity, which results in the lack of exploration and may hinder the learning performance particularly in high-dimensional action space and highly dynamic environment as in our case. To compensate for this problem, The most common way is to add random noise to the outputted continuous action $\tilde{a}^{t}$:
\begin{equation}
a^{t} = f_{a}(\tilde{a}^{t} + \mathcal{N}).
\end{equation}
Here $\mathcal{N}$ can be chosen to suit the environment and $f_{a}$ is a function that maps the continuous action to a binary one. In this paper, we use an Ornstein-Uhlenbeck process \cite{Infeld1945On} in the Brown motion theory to generate temporally correlated exploration in our base station sleep operation problem. However, due to the Curse of Dimensionality, this random noise-based approach can only provide coarse exploration in a high dimensional action space which can not guarantee exploration efficiency. To aid exploration, we apply an explorer network as shown in Fig. \ref{EN}. The parameters $\tilde{W}$ of the explorer network can be obtained by adding a small disturb $\Delta W$ to the parameters of the current actor network
\begin{equation}
\Delta W = \alpha \cdot rand(-1,1) \cdot W,
\end{equation}
where $\alpha$ is the explorer coefficient, and $rand(-1,1)$ generates a random number between -1 and 1.

At each time slot, the current actor network generates the base station active/sleep mode operation $a_{a}^{t}$, and explorer network outputs another BS active/sleep mode operation $a_{e}^{t}$. Our agent will first estimate the two actions and then select the better one which results in a lower cost. If the action generated by the explorer network is considered better than that of current actor network, the parameters of the current actor network will be updated as
\begin{equation}
W^{\prime} = W + \sigma \tilde{W},
\end{equation}
where $\sigma$ is a decreasing factor. Otherwise, the parameters of the current actor network keep unchanged. Through this kind of exploration, our agent is able to do more effective exploration in action generation process and will further accelerate the convergence in training process.

\subsection{DDPG Training Module}
The actor network parameters $\theta_{\pi}$ and the critic network parameters $\theta_{Q}$ are both trained by the backpropagation algorithm. As the ground truth is unknown beforehand, neither the actor network nor the critic network can be trained in a supervised fashion. To train the actor-critic network efficiently, we adopt DDPG training module, which is recently proposed to be robust and stable \cite{LillicrapHPHETS15}.

Directly implementing Q learning with neural networks has been proved to be unstable in many environments, since the network $Q(s^{t}, a^{t}|\theta_{Q})$ being updated is also used for calculating the target value \cite{LillicrapHPHETS15}. To solve this problem, DDPG framework creates target actor network $\pi (s^{t}|\theta_{\pi}^{T})$ and target critic network $Q (s^{t}, a^{t}|\theta_{Q}^{T})$ respectively, which has exactly the same structure as the actor network and critic network. As the weights of the target networks are updated by having them slowly track the learned networks (e.g., for the actor network, $\theta^{T}_{\pi} \leftarrow \tau \theta_{\pi} + (1 - \tau) \cdot \theta^{T}_{\pi} $), the target values provided by the target networks are constrained to change slowly, which greatly improves the stability of learning.

Meanwhile, we use an experience replay buffer $M_{exp}$ to store the experience tuples $<s, a, g, s^{\prime}>$ where $s^{\prime}$ is the next state after taking action $a$ in state $s$. Previous experimental results show that experience replay has greatly improved the learning performance, which is mainly because it can break up the correlation among experiences in a mini-batch. Here we should note that after benchmark transformation and explorer network in the action decision stage, the cost gap $g$ stored in the experience replay buffer maintains a small variance, which is of significant benefit in the DDPG training process. At training time, the agent fetches samples from the buffer randomly rather than using the current experience as standard temporal difference learning \cite{LillicrapHPHETS15}.

\begin{figure}[!t]
	\centering
	\includegraphics[width=0.85\linewidth]{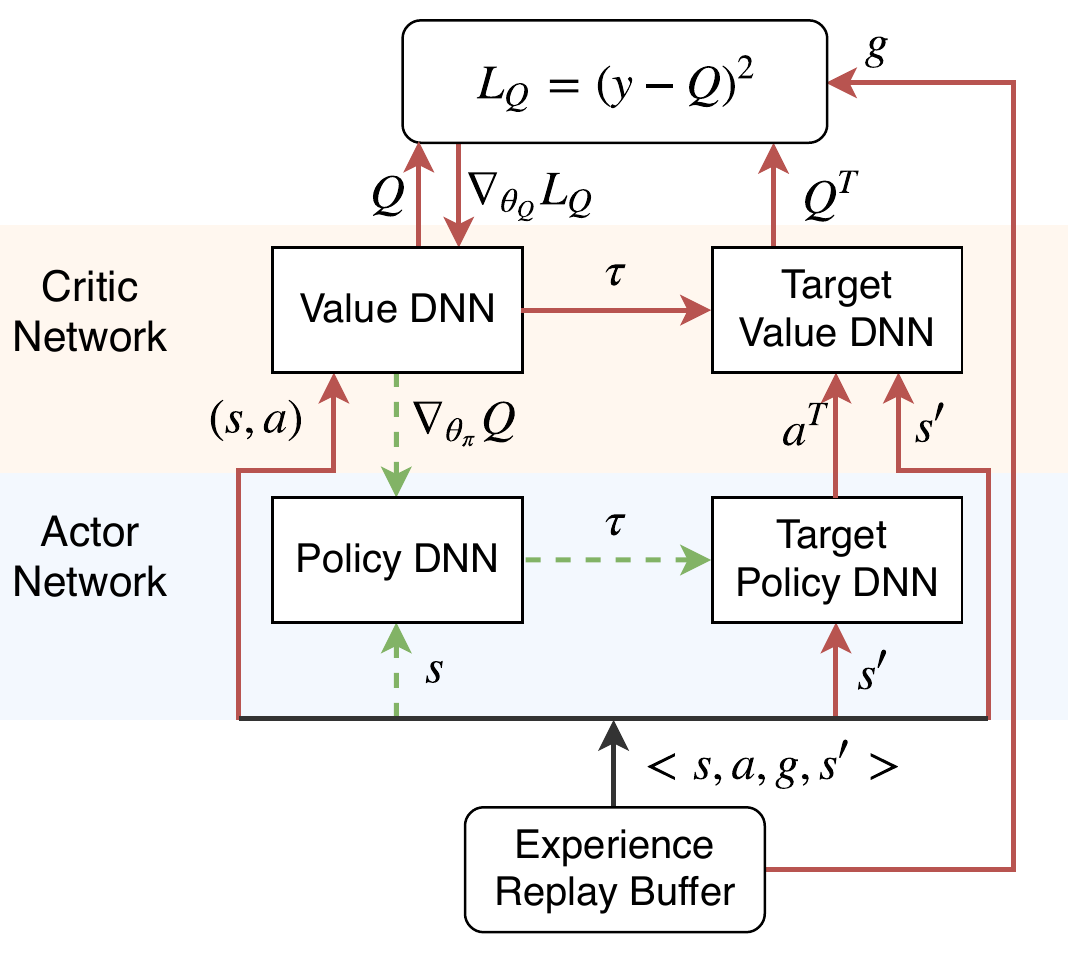}
	\caption{The DDPG procedue for our problem.}
	\label{DDPG}
	\vspace{-10pt}
\end{figure}

As shown in Fig. \ref{DDPG}, given a tuple $<s, a, g, s^{\prime}>$, the critic network estimates the state-action value $Q(s,a|\theta_{Q})$ of the state-action pair $<s, a>$, while the target critic network produces $y$, which is treated as the (approximately) true objective by the critic network. In the training process, we aim to minimize the loss defined as follows
\begin{equation}
L_{Q}(\theta_{Q}) = \mathbb{E}[(y-Q(s,a|\theta_{Q}))^{2}],
\end{equation}
where
\begin{equation}
y = g + \gamma \cdot Q(s^{\prime}, \pi(s^{\prime}|\theta_{\pi}^{T}) | \theta_{Q}^{T}).
\end{equation}
The gradient of $L_{Q}(\theta_{Q})$ with respect to $\theta_{Q}$ can be computed as
\begin{equation}
\label{nable_theta_Q}
\nabla_{\theta_{Q}}L_{Q}(s,a) = \mathbb{E}[2 \cdot (y-Q(s,a|\theta_{Q})) \cdot \nabla_{\theta_{Q}}Q(s,a)].
\end{equation}
Here, $y-Q(s,a|\theta_{Q})$ is the TD-error and $\nabla_{\theta_{Q}}Q(s,a)$ is computed by the chain-rule. In practice, we apply stochastic gradient descent (SGD) method to train the critic network. Specifically, we randomly select a mini-batch experience $<s_{i}, a_{i}, g_{i}, s^{\prime}_{i}>, i = 1, 2, \cdots, N_{e}$,  from the replay buffer to update the critic network parameters $\theta_{Q}$:
\begin{equation}
\label{updateCritic}
\theta_{Q} \gets \theta_{Q} - \frac{\alpha_{Q}}{N_{e}}\sum_{i=1}^{N_{e}}2 \cdot (y_{i}-Q(s_{i},a_{i}|\theta_{Q})) \cdot \nabla_{\theta_{Q}}Q(s_{i},a_{i}),
\end{equation}
where $\alpha_{Q}$ is the learning rate of the critic network and the target value $y_{i}  = g_{i} + \gamma \cdot Q(s_{i}^{\prime}, \pi(s_{i}^{\prime}|\theta_{\pi}^{T}) | \theta_{Q}^{T})$.

\begin{algorithm}[htb]
	\caption{DRL based Base Station Sleep Control}
	\label{algo}
	\begin{algorithmic}[1]
		\Require $N_{exp}$: experience replay buffer maximum size; $N_{e}$: Training batch size;
		\State Randomly initialize the network parameters $\theta_{\pi}$, $\theta_{Q}$, and set $\theta_{\pi}^{T} = \theta_{\pi}$, $\theta_{Q}^{T} = \theta_{Q}$.
		\State $M_{exp} \gets \varnothing$
		\For{each episode $e \in \{1, 2, 3, ...\}$}
		\For{t = 1, 2, ..., T}
		\State /*At the begining of the time slot t*/
		\State Predict the mobile traffic demand $\tilde{D}^{t}$ by GS-STN
		\State Estimate real-time benchmark cost $c^{t}_{base}$
		\State Set current state $s^{t} = (\tilde{D}^{t}, a^{t-1})$
		\State /*At the end of time slot t*/
		\State Generate the action $a^{t}$ and execute it
		\State Observe actual traffic $D^{t}$, cost $c^{t}$ and new state $s^{\prime}$
		\State get the cost gap $g^{t} = c^{t} - c^{t}_{base}$
		\State $M_{exp} \gets M_{exp} \cup \{(s^{t}, a^{t}, g^{t},s^{\prime})\}$
		\If{$|M_{exp}|$ > $N_{exp}$}
		\State Remove the oldest tuple
		\EndIf
		\State Sample a minibatch of $N$ tuples $(s, a, g, s^{\prime})$ from $M_{exp}$
		\State Update critic network parameters with Eqn. (\ref{updateCritic})
		\State Update actor network parameters with Eqn. (\ref{updateActor})
		\State Update target networks:  \begin{equation} \label{eqn2}
		\begin{split}
		\theta_{\pi}^{T} \gets \tau\theta_{\pi} + (1-\tau)\cdot\theta_{\pi}^{T},\\
		\theta_{Q}^{T} \gets \tau\theta_{Q} + (1-\tau)\cdot\theta_{Q}^{T}.
		\end{split}
		\end{equation}
		\EndFor
		\EndFor
	\end{algorithmic}
\end{algorithm}

As for the actor network, we seek to minimize the loss between the optimal action and the action the actor network generates:
\begin{equation}
L_{\pi}(\theta_{\pi}) = \mathbb{E}[(a^{*} - a)^{2}],
\end{equation}
where $a = \pi(s|\theta_{\pi})$ is the generated action by our actor network and $a^{*}$ represents the optimal action. Due to the large action space, $a^{*}$ is almost impossible to obtain. Instead, we update the actor network parameters $\theta_{\pi}$ towards minimizing the Q-value of the outputted action with the deterministic policy gradient \cite{SilverLHDWR14}. To do that, we need the gradient from the critic network with respect to the actor network's output action $\tilde{ a} = \pi(s|\theta_{\pi})$, i.e., $\nabla_{\tilde{a}}Q(s,\tilde{a})$. The complete gradient is
\begin{equation}
\nabla_{\theta_{\pi}}Q = \nabla_{a} Q(s,a|\theta_{Q})|_{a=\pi(s|\theta_{\pi})} \cdot \nabla_{\theta_{\pi}}\pi(s),
\end{equation}
where $\nabla_{\theta_{\pi}}\pi(s)$ is computed by the chain-rule. It has been proved that the stochastic policy gradient, which is the gradient of the policy's performance, is equivalent to the empirical deterministic policy gradient \cite{SilverLHDWR14}.
Thus, in the model training process, we update the actor network parameters $\theta_{\pi}$ with a mini-batch experience just as we do in critic network,
\begin{equation}
\label{updateActor}
\theta_{\pi} \gets \theta_{\pi} - \frac{\alpha_{\pi}}{N_{e}}\sum_{i=1}^{N_{e}} \nabla_{a} Q(s_{i},a|\theta_{Q})|_{a=\pi(s_{i}|\theta_{\pi})} \cdot \nabla_{\theta_{\pi}}\pi(s_{i}),
\end{equation}
where $\alpha_{\pi}$ is the learning rate of actor network.

\subsection{The Holistic Algorithm}
To sum up, the holistic mechanism of our DRL based BS sleep control process is presented in Algorithm \ref{algo}. At the begining of each time slot $t$, the agent first uses the historical mobile traffic data to predict the traffic demand $\tilde{D}^{t}$ for the current time by our proposed GS-STN. Given the current state $s^{t} = (\tilde{D}^{t}, a^{t-1})$ in our environment, the agent will generate the active/sleep mode decisions $a^{t}$ and estimate  the benchmark cost $c_{base}^{t}$ of current state. At the end of time slot $t$, the actual traffic demand and system cost are known, the agent will store the experiences in its replay buffer for future use. And the agent will randomly sample a minibatch of experiences to update the parameters of its networks in order to learn a better policy.

Since the BS sleep control decisions are generated at a large time-scale (e.g., half an hour), the algorithm can be implemented in the cloud datacenter or the central office within a regional network that processes abundant computing resources.

\section{Experiments}
\label{SecExperiments}
\begin{figure}[!t]
	\centering
	\vspace{-10pt}
	\includegraphics[width=0.8\linewidth]{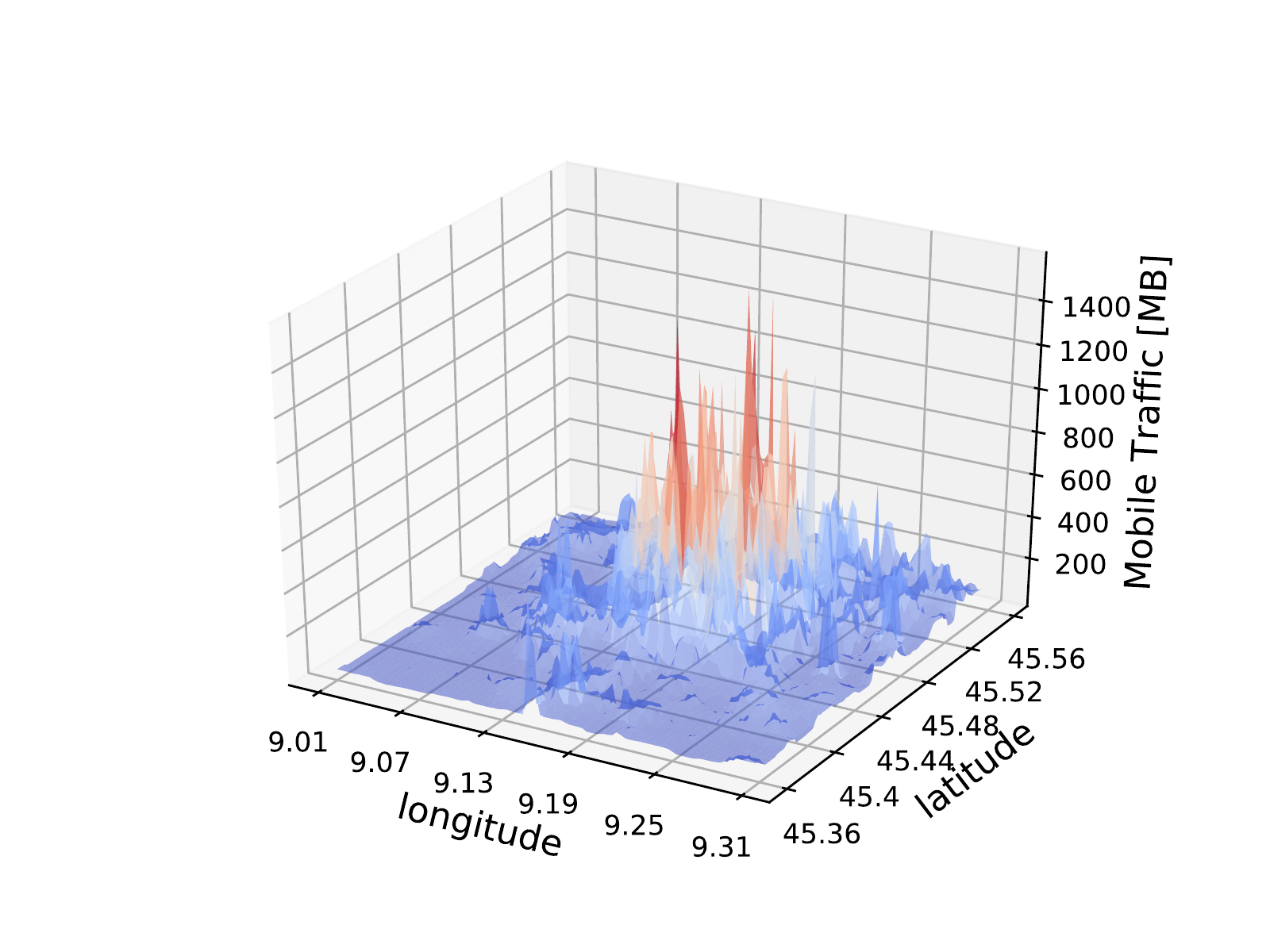}
	\caption{Average mobile data traffic in Milan, Italy. It shows that the mobile traffic demand varies greatly in the geographical space.}
	\label{factor}
	\vspace{-10pt}
\end{figure}
\begin{figure}[!t]
	\centering
	\includegraphics[width=0.8\linewidth]{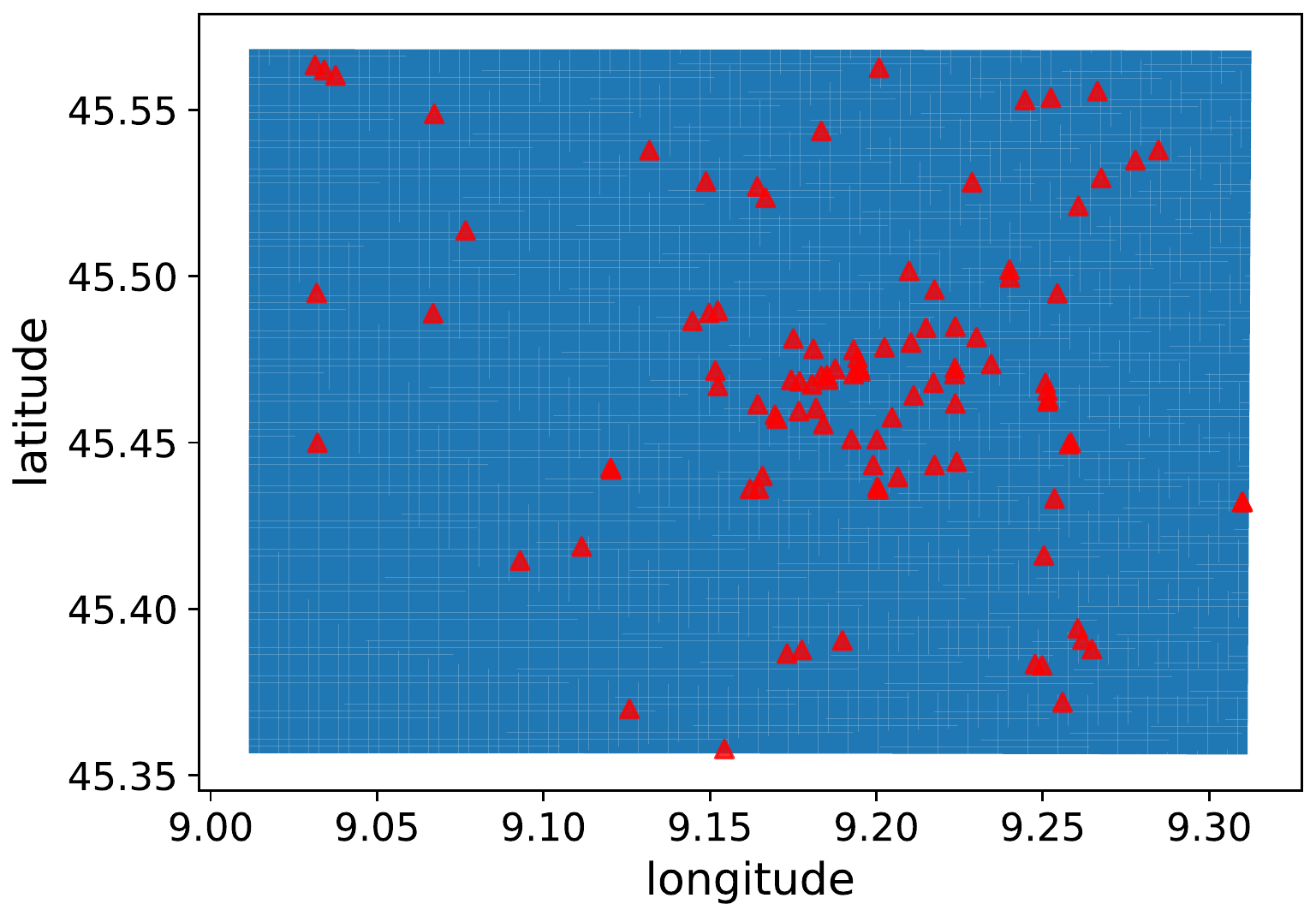}
	\caption{The locations of Base stations in Milan, Italy. BSs are densely deployed in high traffic areas.}
	\label{baseStation}
	\vspace{-10pt}
\end{figure}

\subsection{Dataset Description}
In order to gauge the effectiveness of our proposed method on real networks, we utilize the public available real-world mobile traffic datasets released by Telecom Italia over the city of Milan, Italy in November, 2013 \cite{Barlacchi2015AMD}. To ease the geographical management, the dataset divides the map of Milan into $100\times100$ micro-grids of 250m width. In this paper, we further group them into $10\times10$ grids for ease of implementation. The dataset contains network activity measurement of each grid over 10-minute intervals. As BS model operation requires a slow time scale, we further aggregate into 30-minute time interval. A snapshot of mobile traffic in Milan city is shown in Fig. \ref{factor}. We obtain all the base stations in the same area by utilizing open crowd-sourced dataset from Wigle \cite{RN16}. There are 100 base stations in the city of Milan and the geographical distribution of them is presented in Fig. \ref{baseStation}. According to Fig. \ref{factor} and Fig. \ref{baseStation}, we can find that base stations are densely deployed in high traffic areas.

\subsection{Traffic Forecasting Evaluation}
\subsubsection{\textbf{Hyperparameter Setting}}
In our design, the 2D convolutional layer with filter size $3 \times 3$ and 1D convolutional layer with filter size $1 \times 3$ in GS-STN both contain 10 filters to capture the geographical and semantic similarity, respectively. The number of hidden neurons is 48 in LSTM layer and 100 in dense layer. We set the number of time intervals to 12 (i.e., 6 hours) for LSTM. The activation function is relu for the convolutional layers and sigmoid for the dense layer. These hyperparameters in GS-STN are tuned using the commonly-adopted approach of random search and cross-validation \cite{BergstraB12}. To train the proposed GS-STN model, we use TensorFlow platform and backpropagation algorithm with Adam optimizer, which commonly yields faster convergence compared to traditional stochastic gradient descent method.

\begin{table}[!t]
	\small
	\caption{Traffic forcasting comparison. }
	\label{traffic prediction}       
	\newcommand{\tabincell}[2]{\begin{tabular}{@{}#1@{}}#2\end{tabular}}
	\centering
	\begin{tabular}{|c|c|c|}
		\hline Predicting Methods & NMAE & NRMSE \\
		\hline\hline ARIMA & 0.0545 & 0.1542 \\
		\hline DNN  & 0.0984&  0.1461\\
		\hline CNN-LSTM& 0.0404 & 0.0774 \\
		\hline GS-STN  & \textbf{0.0375}&\textbf{0.0709 }\\
		\hline
	\end{tabular}
	\vspace{-10pt}
\end{table}

\subsubsection{\textbf{Comparing Methods}}
We compare our proposed GS-STN approach with some widely-used methods in traffic forecasting problem. To achieve a fair comparison, all the methods use the historical mobile traffic in the previous 48 time slots for traffic forecasting. For each method, we conduct elaborate hyperparameter tuning to pursue good performance.
\begin{itemize}
	\item \textbf{Auto-Regressive Integrated Moving Average (ARIMA)} : ARIMA is a well-known model which combines moving average and autoregressive components for modeling time series and achieves good performance in time series forecasting tasks \cite{kim2011dynamic}.
	\item \textbf{Deep Neural Network (DNN):} DNN has been identified as a universal function approximator with excellent generalization capability and approximation capacity \cite{HornikSW89}.
	\item \textbf{CNN-LSTM:} This method can capture the spatial-temporal correlations and has been verified to be very efficient for mobile traffic prediction \cite{HuangCL17}.
\end{itemize}
\subsubsection{\textbf{Traffic Forecasting Performance}}
\begin{figure}[!t]
	\centering
	\includegraphics[width=0.85\linewidth]{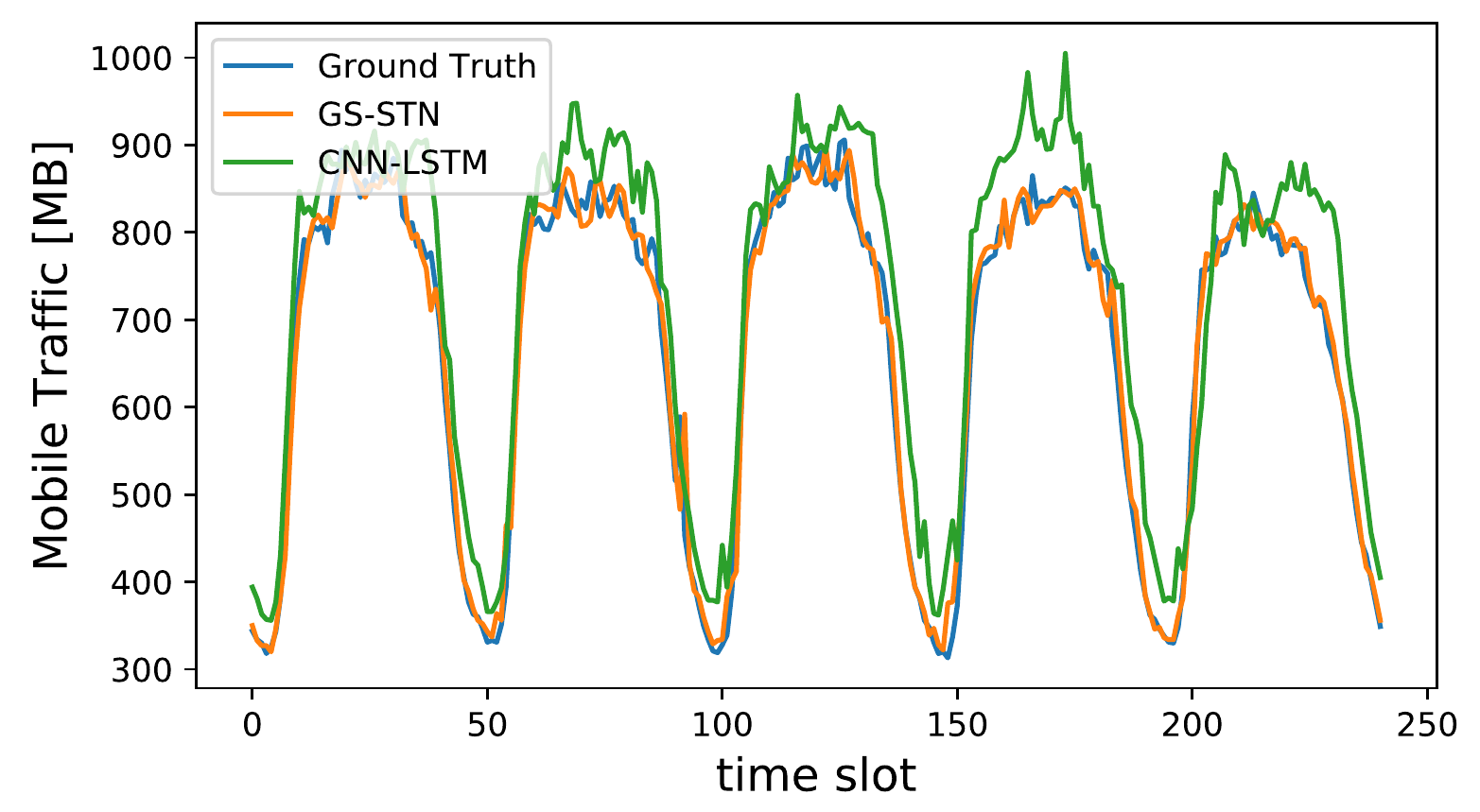}
	\caption{Traffic volume predicted by CNN-LSTM and GS-STN over 240 time slots (5 days), as well as ground truth measurement in grid 64. }
	\label{traffics}
	\vspace{-10pt}
\end{figure}

To quantify the performance of the proposed GS-STN method and existing prediction methods, we use the Normalised Mean Absolute Error (NMAE) and Normalised Root Mean Square Error (NRMSE) \cite{ZhangP18} as given below:
\begin{equation}
NMAE = \frac{1}{\bar{d}}\sum_{k=1}^{N}\frac{|\bar{d_{k}}-d_{k}|}{N},
\end{equation}

\begin{equation}
NRMSE = \frac{1}{\bar{d}}\sqrt{\sum_{k=1}^{N}\frac{(\bar{d_{k}}-d_{k})^{2}}{N}},
\end{equation}
where $\bar{d_{k}}$ is the predicted mobile traffic demand of the geographical grid $k$ and $d_{k}$ is the corresponding ground-truth value. N denotes the total number of grids and $\bar{d}$ is the mean of mobile traffic. Therefore, smaller NMAE and NRMSE mean more accurate traffic predictions.

\begin{figure}[!t]
	\centering
	\includegraphics[width=0.9\linewidth]{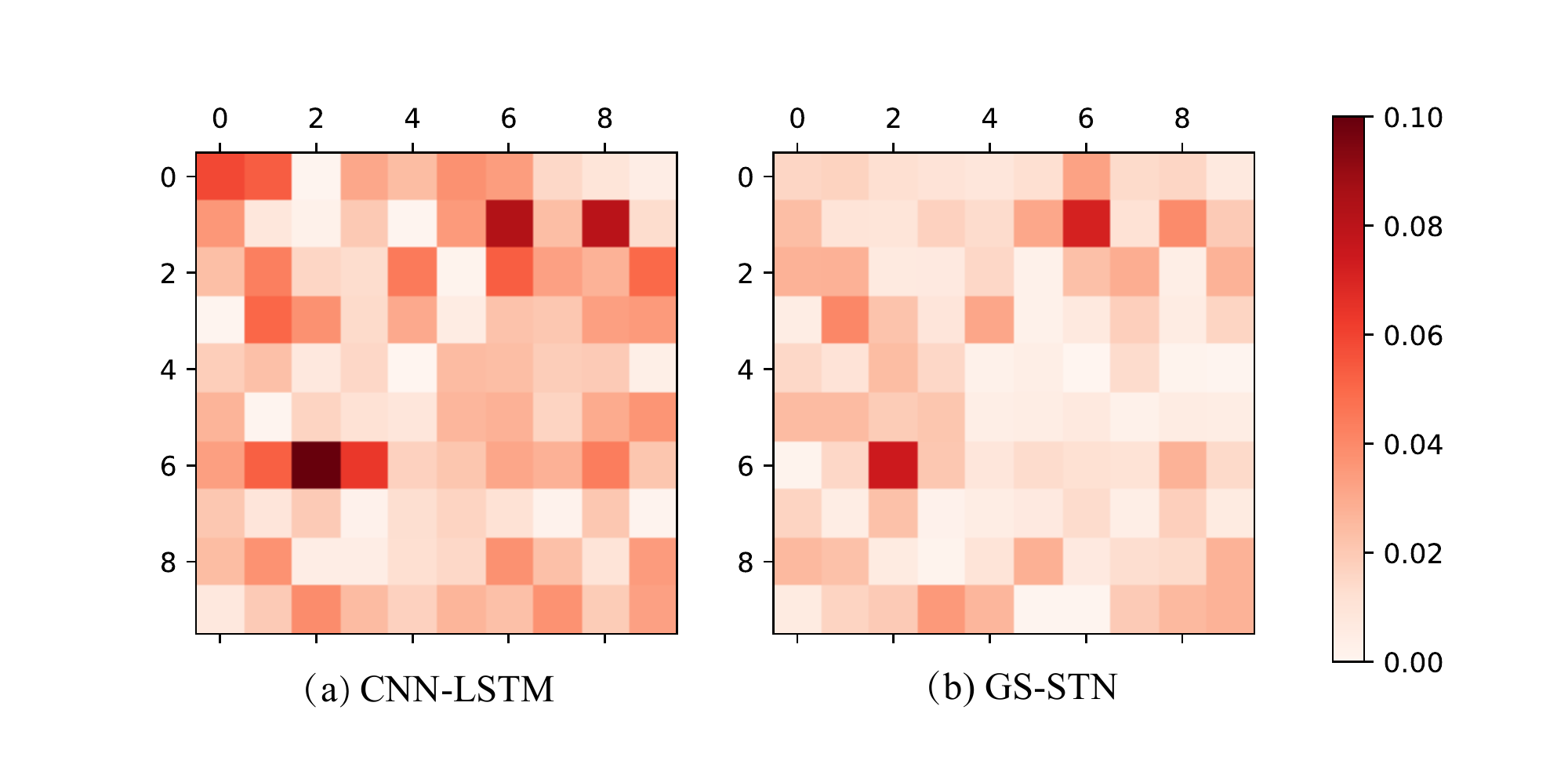}
	\caption{The normalised prediction error of CNN-LSTM and GS-STN.}
	\label{trafficsError}
	\vspace{-10pt}
\end{figure}

Table \ref{traffic prediction} shows the traffic prediction results. We can see that our proposed GS-STN method achieves the smallest NMAE and NRMSE. More specifically, it can achieve $8.4\%$ performance improvement of NRMSE comparing with CNN-LSTM method, which demonstrates that the similarity of mobile traffic demand in semantic space is also important for mobile traffic prediction. To better demonstrate the effectiveness of GS-STN method, we examine the traffic volume predicted by our GS-STN method and the CNN-LSTM approach from both temporal and spatial perspectives. In Fig. \ref{traffics}, we observe that the output of GS-STN follows very closely with the ground truth, while there exists a clear gap between the performance of CNN-LSTM and the ground truth. This is mainly due to the fact that CNN-LSTM utilizes the traffic information of neighboring regions while some of these neighbors may have different traffic patterns with the target region, resulting in inaccurate traffic prediction. While in spatial view, we calculate the normalized prediction error of both CNN-LSTM and GS-STN with the ground truth as shown in Fig. \ref{trafficsError}.  The lighter color stands for more accurate prediction, thus it is obvious that GS-STN can control the prediction error with a certain range of less than $0.06$ and is more effective than CNN-LSTM. The benefit of precise traffic forecasting in our BS sleep control problem will be demonstrated in Section \ref{Sleeping Evaluation}.

\subsection{Base Station Sleep Evaluation}
\label{Sleeping Evaluation}

\subsubsection{\textbf{Hyperparameter Settings}}
In our experiment, the state is 200-dimensional which is composed of the mobile traffic in 100 grids and the operation modes of 100 BSs. Accordingly, the action space is $2^{100}$. We summarize the DNN designs and  key system settings of our DeepBSC framework in Table \ref{DNN} and Table \ref{systemSettings}, respectively. The power parameters in our experiments are from real-network measurements \cite{DeruyckJM14}. Specifically, we train our DeepBSC framework on the dataset during the first 20 days and evaluate on the last 10 days.
\begin{table}[!t]
	\caption{DNN Designs for Actor and Critic.}
	\label{DNN}       
	\newcommand{\tabincell}[2]{\begin{tabular}{@{}#1@{}}#2\end{tabular}}
	\centering
	\begin{tabular}{|c|c|c|c|c|c|c|}
		\hline
		\multirow{2}{*}{}&
		\multicolumn{2}{c|}{Layer 1}&\multicolumn{2}{c|}{Layer 2}&\multicolumn{2}{c|}{Output Layer}\\
		\cline{2-7}
		&Activation&Size&Activation&Size&Activation&Size\\
		\hline
		Actor&BN+relu&800&BN+relu&600&sigmoid&100\\
		\hline
		Critic&BN+relu&800&BN+relu&600&linear&1\\
		\hline
	\end{tabular}
\end{table}

\begin{table}[!t]
	\caption{System Settings and Algorithm Parameters.}
	\label{systemSettings}       
	\newcommand{\tabincell}[2]{\begin{tabular}{@{}#1@{}}#2\end{tabular}}
	\centering
	\begin{tabular}{c|c||c|c}
		\hline $P_{i}^{f}$ & 160W  & $P_{i}^{l}$ & 216W\\
		\hline	$\beta_{d}$ & $50W/s$ & $\beta_{s}^{i}$ & $100Wh/time$\\
		\hline $B$ & $100$ & $N$ & $64$ \\
		\hline $\gamma$ & 0.9 & $\tau$  & $10^{-3}$\\
		\hline $\alpha_{\pi}$& $10^{-4}$ &$\alpha_{Q}$ & $10^{-4}$\\
		\hline
	\end{tabular}
	\vspace{-10pt}
\end{table}
\subsubsection{\textbf{Comparing Methods}} The comparing methods for dynamic base station sleep control problem can be divided into two kinds as follows:

\textbf{Baseline algorithms}:
\begin{itemize}
	\item \textbf{Traffic-oblivious sleep control policy (TO):} The sleep mode of each base station is decided with a probability $p_{sleep}$ based on statistical model/information, no matter how the real-time mobile traffic changes \cite{LiuNX16}.
	\item \textbf{Greedy-off algorithm (GOFF):} Starting from the entire BSs $\mathcal{B}$, GOFF finds a solution by iteratively removing the BS with the lowest turn-off detriment per fixed power consumption \cite{SonKYK11}. Due to combinatorial nature of BS sleep control problem, greedy algorithms are widely adopted in literature.
	\item \textbf{Two-threshold policy (TTP):} In order to balance the trade-off between energy saving from sleeping and QoS degradation, TTP designs an active threshold and a sleep threshold to decide whether the action is to be active, to sleep or to stay at the current operation mode. The thresholds not only depend on the traffic volume, but also the QoS degradation cost and BS mode switching cost. TTP is widely-used in literature such as \cite{Kamitsos2010Optimal, Heyman1968, leng2017wait}.
	\item \textbf{Deep Q-Network (DQN)} uses a deep neural network as a function approximator to estimate the optimal Q-value function and has been successfully used in BS sleep control problem \cite{LiuKZN18}.
	\item \textbf{Deep Deterministic Policy Gradient (DDPG)}: Compared with DQN, DDPG can be applied to high-dimensional action space by using the continuous approximation \cite{LillicrapHPHETS15}.
\end{itemize}

\textbf{Variations of our model}:
\begin{itemize}
	\item \textbf{DDPG $+$ BT:} To address the problem that cost estimation in DDPG exhibits a large variance, which is caused by the mobile traffic demand fluctuation, we propose the benchmark transformation (BT) method in cost estimation.
	\item \textbf{DDPG $+$ EN:} As the DDPG directly outputs an action, resulting in the lack of exploration, we add an explorer network (EN) in our model.
	\item \textbf{DDPG $+$ BT $+$ EN:} Both benchmark transformation (BT) strategy and explorer network (EN) are utilized in the DDPG learning process.
\end{itemize}

\subsubsection{\textbf{Performance Evaluation}}
\begin{figure}[!t]
	\centering
	\includegraphics[width=0.8\linewidth]{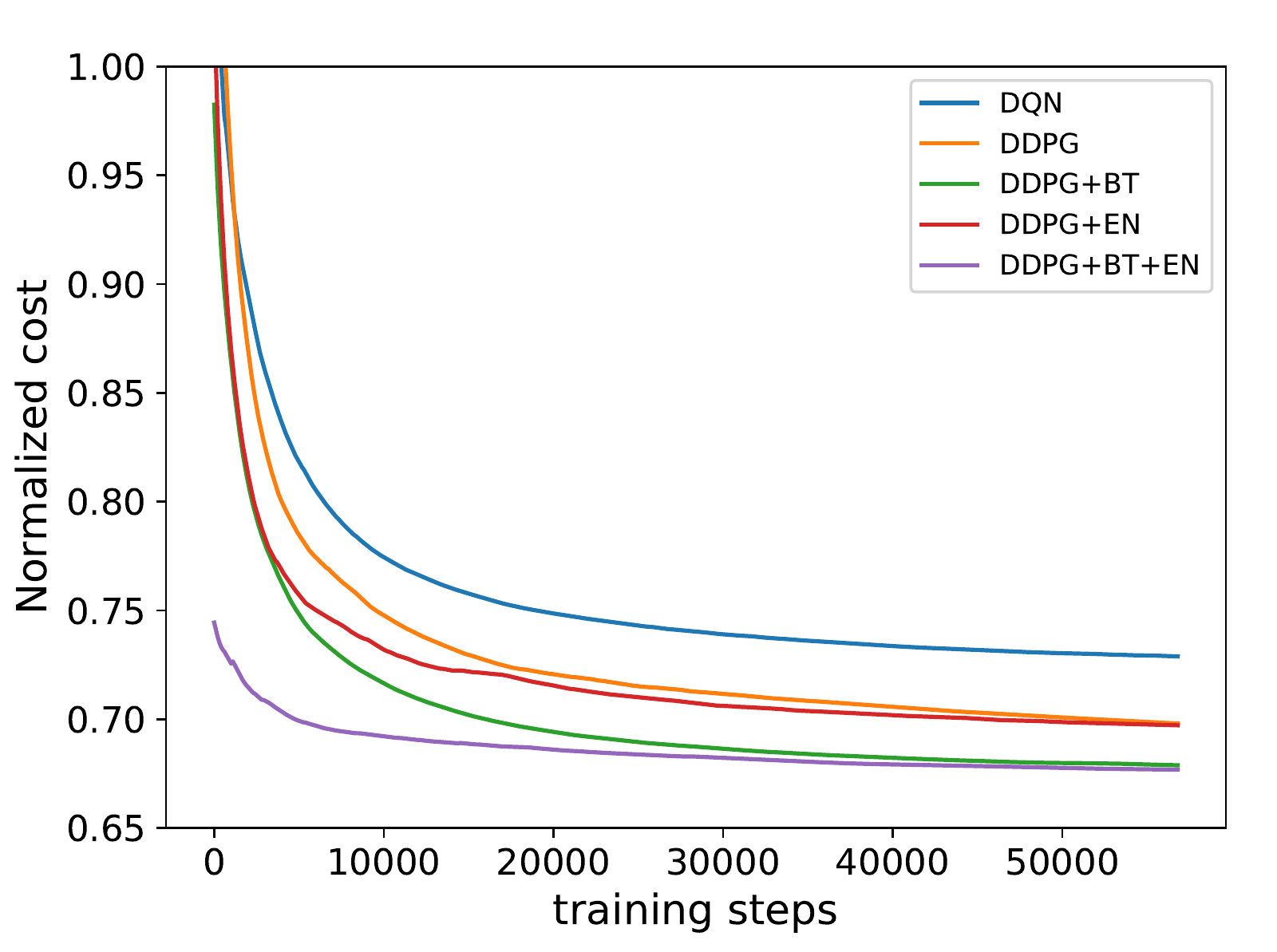}
	\vspace{-10pt}
	\caption{The convergence process of different methods. Our proposed DDPG+BT+EN approach can quickly converge to a very low cost comparing with other methods.}
	\label{convergence}
	\vspace{-10pt}
\end{figure}
We first show the convergence performance of our proposed method comparing with DRL-based baseline models. As illustrated in Fig. \ref{convergence}, DDPG based methods can all converge faster and get lower normalized cost (at least $6\%$ cost reduction) than DQN in the training process. It indicates that DDPG can learn a good policy in our BS sleep control problem which has a large state-action space. When we use benchmark transformation strategy in cost estimation, both DDPG+BT and DDPG+BT+EN methods can converge to a very low cost with $4.464\%$ cost reduction comparing with DDPG approach, which shows that benchmark transformation strategy can effectively handle the large variance problem in cost estimation caused by mobile traffic fluctuation. Moreover, with the explorer network which is added to aid exploration, both DDPG+EN and DDPG+BT+EN can achieve rapid convergence than the case without explorer network. Note that both DQN and DDPG are strong baselines, which are widely used as state-of-the-art methods, so the performance improvement of our proposed approach is significant.

\begin{figure}[!t]
	\centering
	\includegraphics[width=0.85\linewidth]{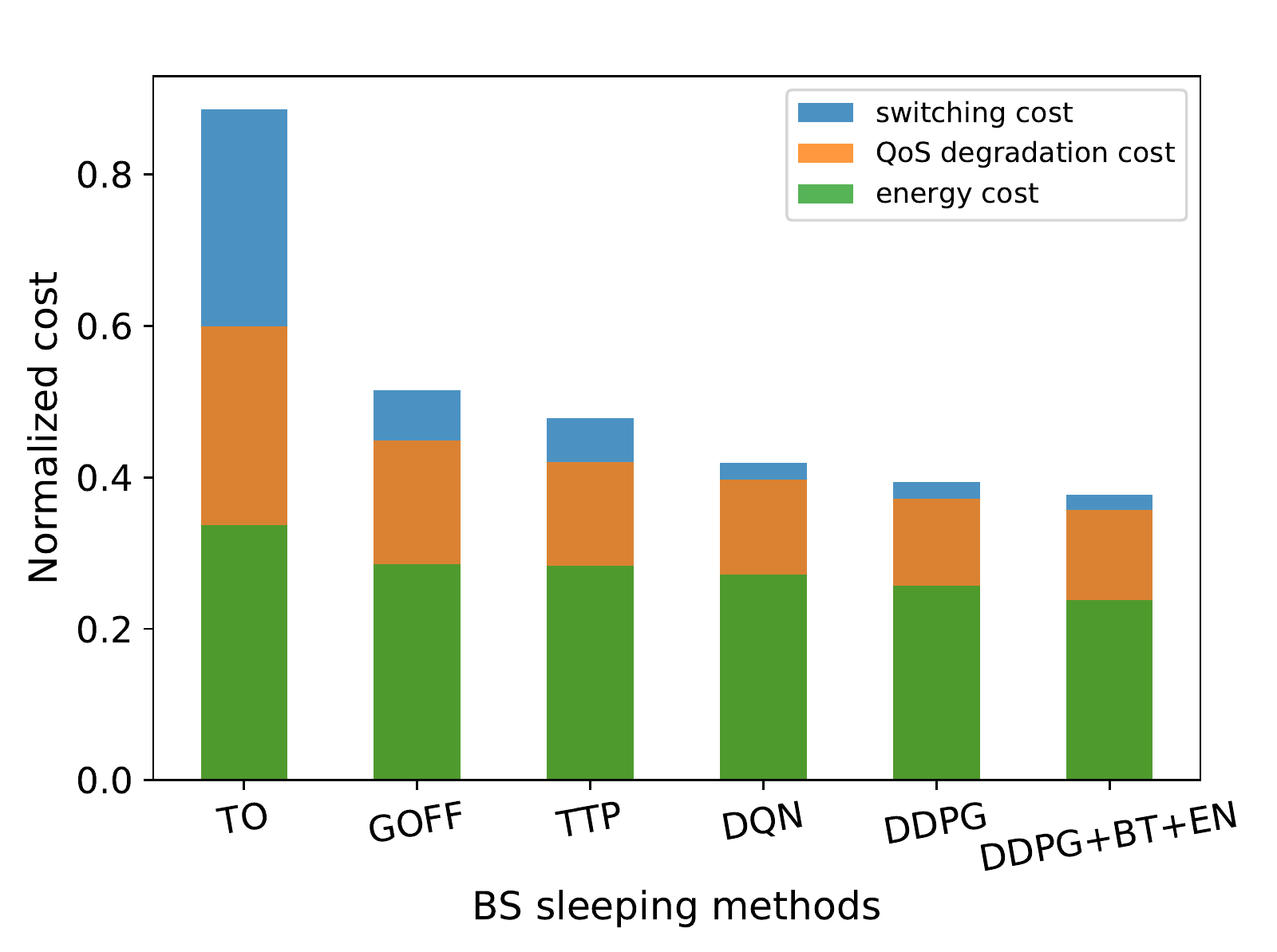}
	\vspace{-10pt}
	\caption{The normalized system cost obtained by different kinds of BS sleep control methods.}
	\label{diffMethods}
	\vspace{-10pt}
\end{figure}

Fig. \ref{diffMethods} illustrates the normalized system cost under different BS sleep control methods. We can see that there is an obvious cost reduction when using GOFF method to deactivate BSs with lower energy efficiency comparing with traffic-oblivious sleep control methods. The widely-used two-threshold policy can maintain a lower system cost  than GOFF method. When we use DRL based approaches, the system cost will be further reduced. Specifically, our proposed DDPG+BT+EN method has a cost reduction of $7.18\%$ than DDPG method, $12.8\%$ than DQN method, $21.19\%$ than TTP method, $29.08\%$ than GOFF method and $57.33\%$ than traffic-oblivious method, which is significant for the energy saving issue in large-scale cellular networks. Moreover, each kind of the cost (e.g., switching cost, energy cost) produced by our DDPG+BT+EN method is the lowest, which indicates the high energy efficiency of our proposed method.

\begin{figure}[!t]
	\centering
	\includegraphics[width=0.8\linewidth]{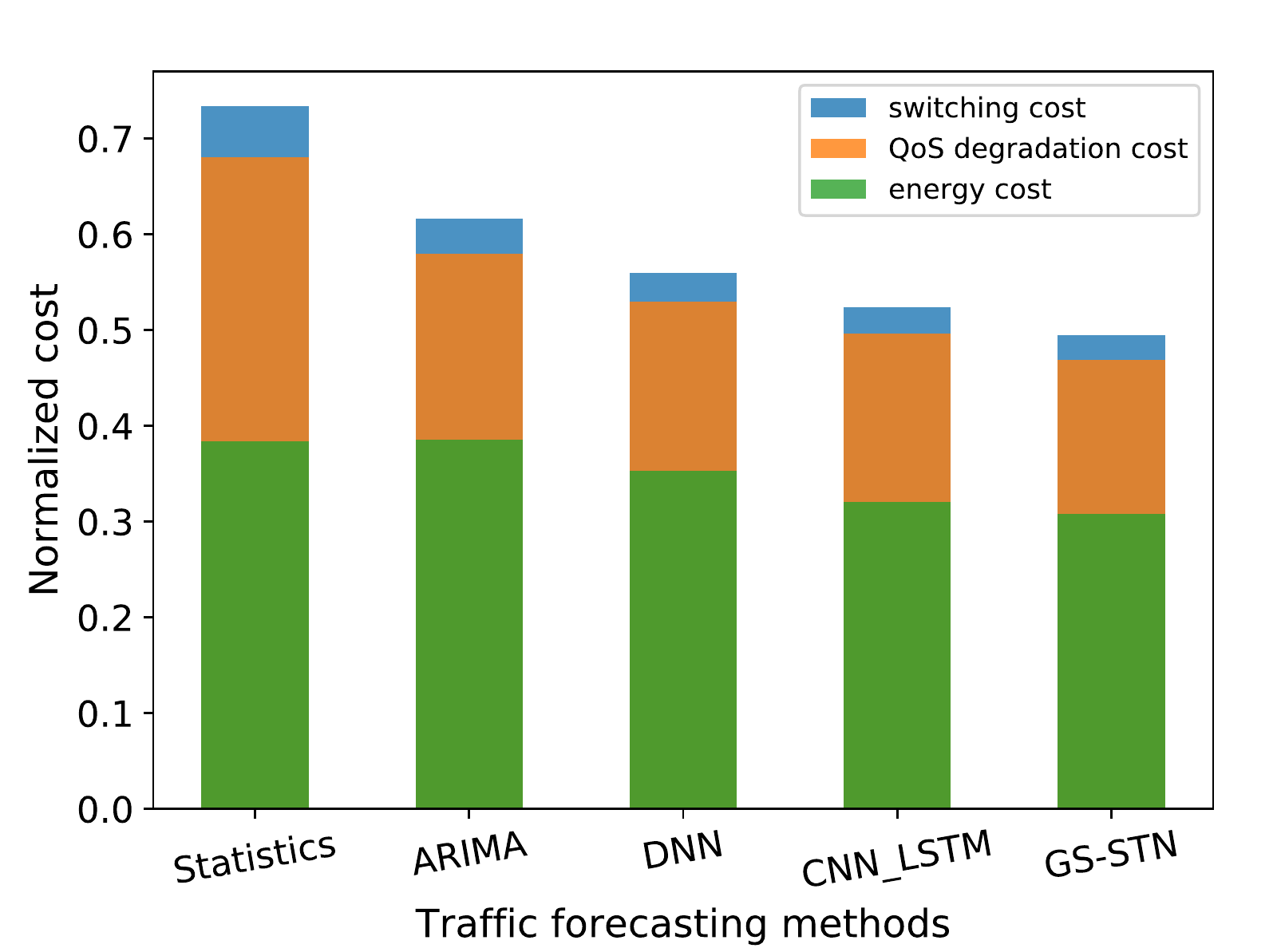}
	\vspace{-5pt}
	\caption{The normalized system cost when using different traffic forecasting methods.}
	\label{diffTraffic}
	\vspace{-10pt}
\end{figure}

\begin{figure}[!t]
	\centering
	\subfigure[Active BS size]{
		\includegraphics[width=0.45\linewidth]{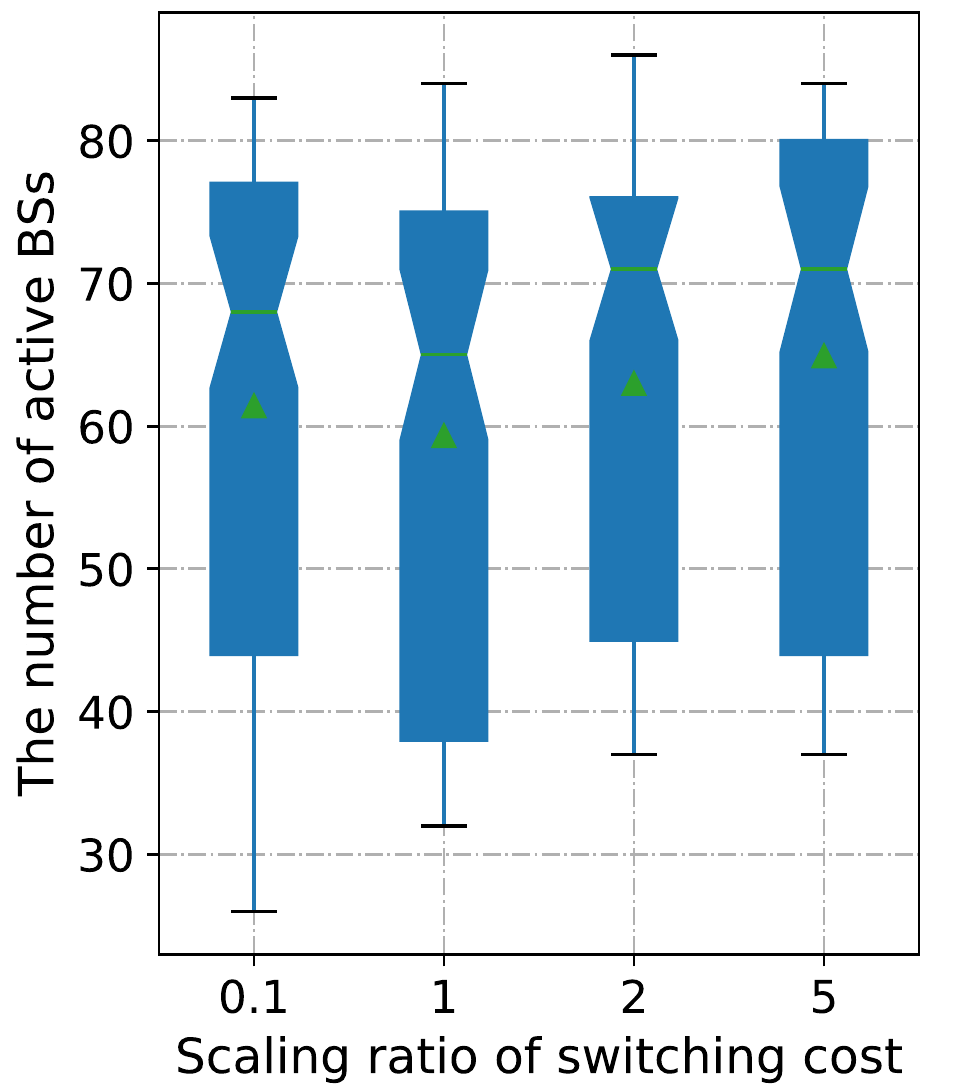}
	}
	\subfigure[System cost]{
		\includegraphics[width=0.45\linewidth]{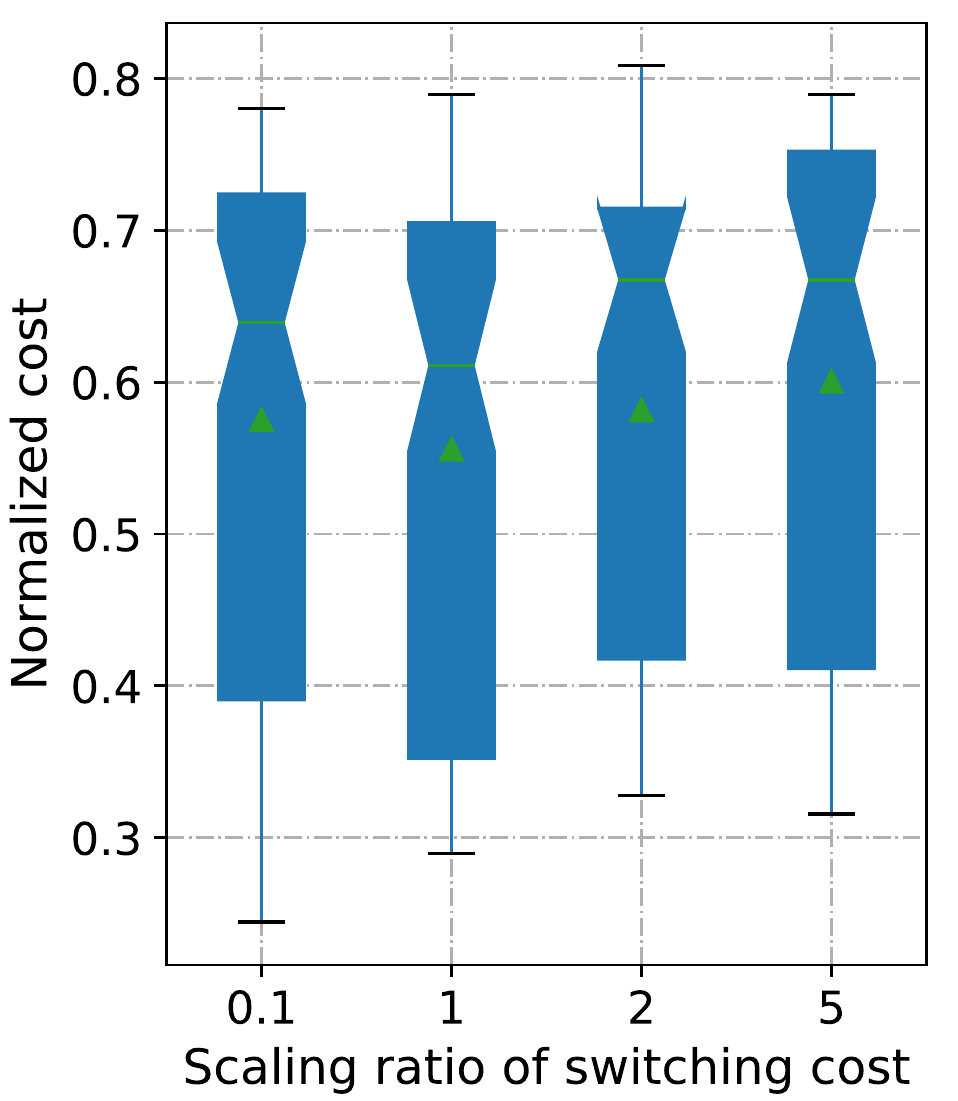}
	} 	
	\vspace{-5pt}
	\caption{The effect of the switching cost on active BS size and total system cost.}
	\label{box}
	\vspace{-17pt}
\end{figure}

To illustrate the impact of the traffic model on energy saving in our BS sleep control problem, we consider a scheme without prediction as the baseline. Here we use the statistical traffic information and consider that the mobile traffic in each grid is given by its statistical mean. As depicted in Fig. \ref{diffTraffic}, with predicted mobile traffic demand, our proposed BS sleep control approach can make better active/sleep mode decisions, thus leading to lower system cost comparing with statistic traffic pattern which has no knowledge of the ever-changing mobile traffic demand. Moreover, the better we can predict the model traffic demand, the better we can pre-activate BSs to meet the demand and avoid unnecessary energy consumption. For example, with the traffic demand predicted by GS-STN, we can get $5.81\%$  and $20.53\%$ system cost reduction than the traffic demand obtained from CNN-LSTM and ARIMA, respectively.

We also examine the effect of the BS switching cost on our BS sleep control problem, by multiplying the switching cost of each BS with various scaling ratios. We adopt boxplot to graphically depict the five-number summary of the number of active BSs, which consists of the smallest observation, lower quartile, median, upper quartile and largest observation. The upper quartile, median and lower quartile make up a box with compartments, which is the blue areas in Fig. \ref{box}. The spacings between different parts of the box can help us indicate the variance and skew in the data distribution. We also show the mean of active BS number with the green triangles. As illustrated in Fig. \ref{box}, when the ratio of switching cost is $0.1$, which implies we are not very concerned with switching cost, the modes of BSs will change frequently and dramatically. In our case, the number of active BSs can range from $26$ to $83$. When the ratio of switching cost is $1$, our system will determine the modes of BSs wisely by considering both switching cost and traffic demand fluctuation, without getting into extreme situations (e.g., frequent or redundant BS activation). Thus, the average number of active BSs in the long run is the lowest. As a result, it brings the lowest total system cost as depicted in Fig. \ref{box} (b). When the ratio of switching cost goes high, our system will tend to keep more BSs active all the time for fear of the high switching cost, leading to a high system cost.

\begin{figure}[!t]
	\centering
	\vspace{-10pt}
	\includegraphics[width=0.85\linewidth]{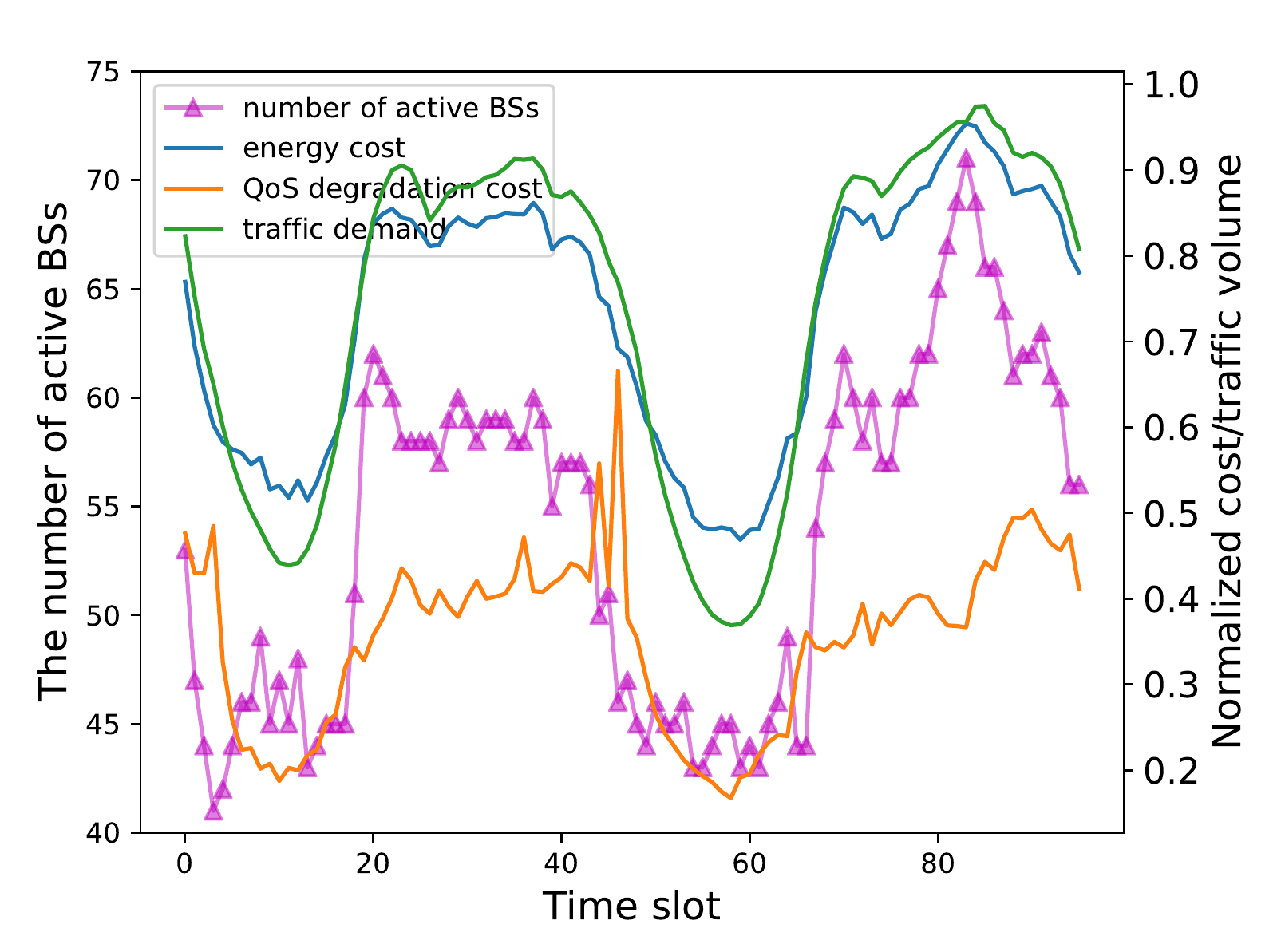}
	\vspace{-5pt}
	\caption{The number of active BSs, corresponding with different kinds of costs in a time span of two days produced by DDPG.}
	\label{nobench}
	\vspace{-15pt}
\end{figure}

\begin{figure}[!t]
	\centering
	\includegraphics[width=0.85\linewidth]{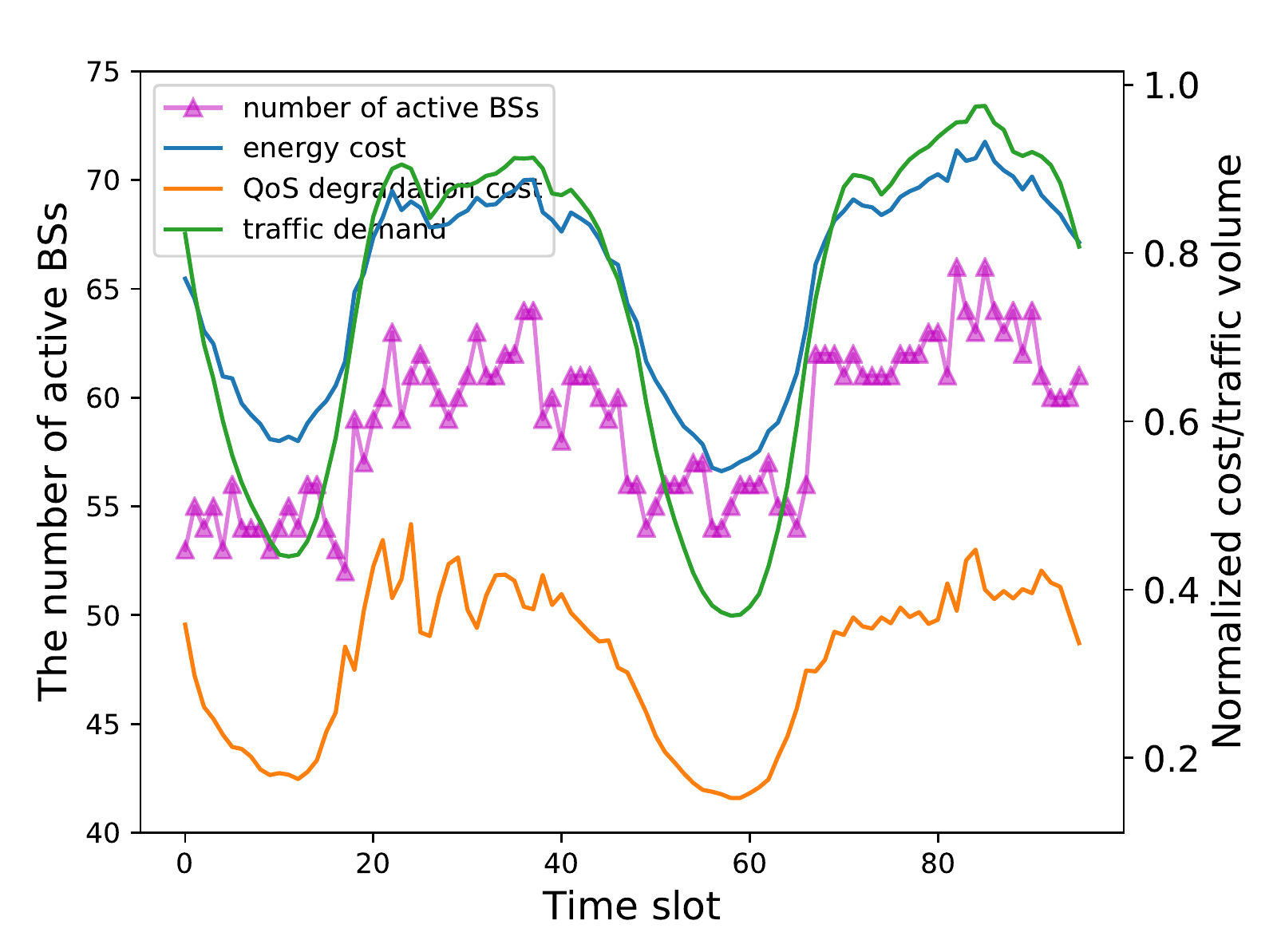}
	\vspace{-5pt}
	\caption{The number of active BSs, corresponding with different kinds of costs in a time span of two days produced by DDPG+BT+EN.}
	\label{DDPG+reduction}
	\vspace{-10pt}
\end{figure}

To further investigate how our proposed approach can achieve lower system cost, we provide the BS active/sleep decision making results together with corresponding costs in two days generated by DDPG and DDPG+BT+EN approaches, as shown in Fig. \ref{nobench} and Fig. \ref{DDPG+reduction}. We can see that in the two figures, the number of active BSs changes as the mobile traffic fluctuates, which in turn causes the changes of different kinds of system costs (e.g., energy cost, QoS degradation cost). However, in the environment with the same mobile traffic demand fluctuation, the two approaches make significantly different sequential decisions. In Fig. \ref{nobench}, the number of active BSs changes dramatically, and thus there may exist the case that user's QoS declines rapidly according to terrible BS mode control. For example, at time slot 45, user's QoS degradation cost has a sharp rise as the DDPG agent deactivates some BSs unwisely. This is because that the large variance in cost estimation caused by the environment changes degrades the performance of DDPG in the training process. On the contrary, DDPG+BT+EN  approach eliminates the impact of varying traffic demand on cost estimation and strengthens the exploration in the training process, leading to a wiser sequential decision making strategy. As depicted in Fig. \ref{DDPG+reduction}, the BS active/sleep decision made by DDPG+BT+EN ranges slightly, which can thus maintain a stable QoS as well as lower system cost.

\section{Conclusion}
\label{SecConclusion}
In this paper we design a traffic-aware DRL-based dynamic BS sleep control framework, named DeepBSC, for energy saving in large-scale cellular network. We first develop a GS-STN network for precise mobile traffic forecasting, which is crucial for BS sleep control. Then we formulate the BS sleep control problem as an MDP to minimize the long-term energy consumption with the users' QoS and switching cost considered. To solve the MDP, we adopt Actor-Critic reinforcement learning architecture and both actor and critic network are approximated by DNN. We propose benchmark transformation strategy to alleviate the large fluctuation in cost estimation caused by the highly fluctuating traffic loads and devise an explorer network to aid exploration to significantly enhance the learning performance. By extensive experiments with real-world large-scale cellular network dataset, we demonstrate the effectiveness of our proposed DeepBSC framework. We hope that this work can help to elicit escalating attention on large-scale data-driven approach for future green cellular network design by leveraging the power of AI.

\ifCLASSOPTIONcaptionsoff
\newpage
\fi

\bibliographystyle{IEEEtran}
\bibliography{reference}

\begin{thebibliography}{10}
\providecommand{\url}[1]{#1}
\csname url@samestyle\endcsname
\providecommand{\newblock}{\relax}
\providecommand{\bibinfo}[2]{#2}
\providecommand{\BIBentrySTDinterwordspacing}{\spaceskip=0pt\relax}
\providecommand{\BIBentryALTinterwordstretchfactor}{4}
\providecommand{\BIBentryALTinterwordspacing}{\spaceskip=\fontdimen2\font plus
\BIBentryALTinterwordstretchfactor\fontdimen3\font minus
  \fontdimen4\font\relax}
\providecommand{\BIBforeignlanguage}[2]{{%
\expandafter\ifx\csname l@#1\endcsname\relax
\typeout{** WARNING: IEEEtran.bst: No hyphenation pattern has been}%
\typeout{** loaded for the language `#1'. Using the pattern for}%
\typeout{** the default language instead.}%
\else
\language=\csname l@#1\endcsname
\fi
#2}}
\providecommand{\BIBdecl}{\relax}
\BIBdecl

\bibitem{marsan2009optimal}
M.~A. Marsan, L.~Chiaraviglio, D.~Ciullo, and M.~Meo, ``Optimal energy savings
  in cellular access networks,'' in \emph{2009 IEEE International Conference on
  Communications Workshops}, 2009, pp. 1--5.

\bibitem{richter2009energy}
F.~Richter, A.~J. Fehske, and G.~P. Fettweis, ``Energy efficiency aspects of
  base station deployment strategies for cellular networks,'' in \emph{IEEE
  70th Vehicular Technology Conference Fall}, 2009, pp. 1--5.

\bibitem{ZhengCCLZ15}
J.~Zheng, Y.~Cai, X.~Chen, R.~Li, and H.~Zhang, ``Optimal base station sleeping
  in green cellular networks: A distributed cooperative framework based on game
  theory,'' \emph{IEEE Transactions on Wireless Communications}, vol.~14,
  no.~8, pp. 4391--4406, 2015.

\bibitem{8876870}
E.~{Li}, L.~{Zeng}, Z.~{Zhou}, and X.~{Chen}, ``Edge ai: On-demand accelerating
  deep neural network inference via edge computing,'' \emph{IEEE Transactions
  on Wireless Communications}, vol.~19, no.~1, pp. 447--457, 2020.

\bibitem{8463562}
T.~{Ouyang}, Z.~{Zhou}, and X.~{Chen}, ``Follow me at the edge: Mobility-aware
  dynamic service placement for mobile edge computing,'' \emph{IEEE Journal on
  Selected Areas in Communications}, vol.~36, no.~10, pp. 2333--2345, 2018.

\bibitem{WuZN13}
J.~Wu, S.~Zhou, and Z.~Niu, ``Traffic-aware base station sleeping control and
  power matching for energy-delay tradeoffs in green cellular networks,''
  \emph{IEEE Transactions on Wireless Communications}, vol.~12, no.~8, pp.
  4196--4209, 2013.

\bibitem{Kamitsos2010Optimal}
I.~Kamitsos, L.~Andrew, H.~Kim, and M.~Chiang, ``Optimal sleep patterns for
  serving delay-tolerant jobs,'' in \emph{Proceedings of the 1st International
  Conference on Energy-Efficient Computing and Networking}.\hskip 1em plus
  0.5em minus 0.4em\relax ACM, 2010, pp. 31--40.

\bibitem{Heyman1968}
D.~P. Heyman, ``Optimal operating policies for m/g/1 queuing systems,''
  \emph{Operations Research}, vol.~16, no.~2, pp. 362--382, 1968.

\bibitem{KimLCCL11}
H.-W. Kim, J.-H. Lee, Y.-H. Choi, Y.-U. Chung, and H.~Lee, ``Dynamic bandwidth
  provisioning using arima-based traffic forecasting for mobile wimax,''
  \emph{Computer Communications}, vol.~34, no.~1, pp. 99--106, 2011.

\bibitem{ZhangP18}
C.~Zhang and P.~Patras, ``Long-term mobile traffic forecasting using deep
  spatio-temporal neural networks,'' in \emph{Proceedings of the Eighteenth ACM
  International Symposium on Mobile Ad Hoc Networking and Computing}.\hskip 1em
  plus 0.5em minus 0.4em\relax ACM, 2018, pp. 231--240.

\bibitem{LiuKZN18}
J.~Liu, B.~Krishnamachari, S.~Zhou, and Z.~Niu, ``Deepnap: Data-driven base
  station sleeping operations through deep reinforcement learning,'' \emph{IEEE
  Internet of Things Journal}, vol.~5, no.~6, pp. 4273--4282, 2018.

\bibitem{SonKYK11}
K.~Son, H.~Kim, Y.~Yi, and B.~Krishnamachari, ``Base station operation and user
  association mechanisms for energy-delay tradeoffs in green cellular
  networks,'' \emph{IEEE journal on selected areas in communications}, vol.~29,
  no.~8, pp. 1525--1536, 2011.

\bibitem{SuttonB98}
R.~S. Sutton and A.~G. Barto, ``Reinforcement learning: An introduction,''
  \emph{{IEEE} Trans. Neural Networks}, vol.~9, no.~5, pp. 1054--1054, 1998.

\bibitem{OhK10}
E.~Oh and B.~Krishnamachari, ``Energy savings through dynamic base station
  switching in cellular wireless access networks,'' in \emph{IEEE Global
  Telecommunications Conference GLOBECOM}, 2010, pp. 1--5.

\bibitem{zhou2009green}
S.~Zhou, J.~Gong, Z.~Yang, Z.~Niu, and P.~Yang, ``Green mobile access network
  with dynamic base station energy saving,'' in \emph{ACM MobiCom}, vol.~9, no.
  262, 2009, pp. 10--12.

\bibitem{Tikunov2007Traffic}
D.~Tikunov and T.~Nishimura, ``Traffic prediction for mobile network using
  holt-winter's exponential smoothing,'' in \emph{15th International Conference
  on Software, Telecommunications and Computer Networks}.\hskip 1em plus 0.5em
  minus 0.4em\relax IEEE, 2007, pp. 1--5.

\bibitem{HuangCL17}
C.~Huang, C.~Chiang, and Q.~Li, ``A study of deep learning networks on mobile
  traffic forecasting,'' in \emph{28th {IEEE} Annual International Symposium on
  Personal, Indoor, and Mobile Radio Communications, {PIMRC}}, 2017, pp. 1--6.

\bibitem{WongYP12}
W.-T. Wong, Y.-J. Yu, and A.-C. Pang, ``Decentralized energy-efficient base
  station operation for green cellular networks,'' in \emph{IEEE Global
  Communications Conference (GLOBECOM)}, 2012, pp. 5194--5200.

\bibitem{LiaoHLL14}
W.-C. Liao, M.~Hong, Y.-F. Liu, and Z.-Q. Luo, ``Base station activation and
  linear transceiver design for optimal resource management in heterogeneous
  networks,'' \emph{IEEE Transactions on Signal Processing}, vol.~62, no.~15,
  pp. 3939--3952, 2014.

\bibitem{ZhuangGH16}
B.~Zhuang, D.~Guo, and M.~L. Honig, ``Energy-efficient cell activation, user
  association, and spectrum allocation in heterogeneous networks,'' \emph{IEEE
  Journal on Selected Areas in Communications}, vol.~34, no.~4, pp. 823--831,
  2016.

\bibitem{leng2017wait}
B.~Leng, X.~Guo, X.~Zheng, B.~Krishnamachari, and Z.~Niu, ``A wait-and-see
  two-threshold optimal sleeping policy for a single server with bursty
  traffic,'' \emph{IEEE Transactions on Green Communications and Networking},
  vol.~1, no.~4, pp. 528--540, 2017.

\bibitem{LiZCPZ14}
R.~Li, Z.~Zhao, X.~Chen, J.~Palicot, and H.~Zhang, ``Tact: A transfer
  actor-critic learning framework for energy saving in cellular radio access
  networks,'' \emph{IEEE transactions on wireless communications}, vol.~13,
  no.~4, pp. 2000--2011, 2014.

\bibitem{LiZWZZ12}
R.~Li, Z.~Zhao, Y.~Wei, X.~Zhou, and H.~Zhang, ``Gm-pab: a grid-based energy
  saving scheme with predicted traffic load guidance for cellular networks,''
  in \emph{IEEE International Conference on Communications (ICC)}, 2012, pp.
  1160--1164.

\bibitem{KangSZ12}
T.~Kang, X.~Sun, and T.~Zhang, ``Base station switching based dynamic energy
  saving algorithm for cellular networks,'' in \emph{3rd {IEEE} International
  Conference on Network Infrastructure and Digital Content, {IC-NIDC}}, 2012,
  pp. 66--70.

\bibitem{newell2013applications}
C.~Newell, \emph{Applications of queueing theory}.\hskip 1em plus 0.5em minus
  0.4em\relax Springer Science \& Business Media, 2013, vol.~4.

\bibitem{DBLP:journals/ras/Krose95}
B.~J.~A. Kr{\"{o}}se, ``Learning from delayed rewards,'' \emph{Robotics and
  Autonomous Systems}, vol.~15, no.~4, pp. 233--235, 1995.

\bibitem{XuLWZJ17}
H.~Wang, F.~Xu, Y.~Li, P.~Zhang, and D.~Jin, ``Understanding mobile traffic
  patterns of large scale cellular towers in urban environment,'' in
  \emph{Proceedings of the 2015 Internet Measurement Conference}, 2015, pp.
  225--238.

\bibitem{FumoFS17}
A.~Furno, M.~Fiore, and R.~Stanica, ``Joint spatial and temporal classification
  of mobile traffic demands,'' in \emph{IEEE INFOCOM 2017-IEEE Conference on
  Computer Communications}, 2017, pp. 1--9.

\bibitem{Yao0KTJLGYL18}
H.~Yao, F.~Wu, J.~Ke, X.~Tang, Y.~Jia, S.~Lu, P.~Gong, J.~Ye, and Z.~Li, ``Deep
  multi-view spatial-temporal network for taxi demand prediction,'' in
  \emph{Proceedings of the Thirty-Second {AAAI} Conference on Artificial
  Intelligence, (AAAI-18)}, 2018, pp. 2588--2595.

\bibitem{HochreiterS97}
S.~Hochreiter and J.~Schmidhuber, ``Long short-term memory,'' \emph{Neural
  computation}, vol.~9, no.~8, pp. 1735--1780, 1997.

\bibitem{Informatik2001Gradient}
S.~Hochreiter, Y.~Bengio, P.~Frasconi, J.~Schmidhuber \emph{et~al.},
  \emph{Gradient flow in recurrent nets: the difficulty of learning long-term
  dependencies}.\hskip 1em plus 0.5em minus 0.4em\relax A field guide to
  dynamical recurrent neural networks. IEEE Press, 2001.

\bibitem{ye2018drag}
J.~Ye and Y.-J.~A. Zhang, ``Drag: Deep reinforcement learning based base
  station activation in heterogeneous networks,'' \emph{arXiv preprint
  arXiv:1809.02159}, 2018.

\bibitem{WeiYSH18}
Y.~Wei, F.~R. Yu, M.~Song, and Z.~Han, ``User scheduling and resource
  allocation in hetnets with hybrid energy supply: An actor-critic
  reinforcement learning approach,'' \emph{IEEE Transactions on Wireless
  Communications}, vol.~17, no.~1, pp. 680--692, 2017.

\bibitem{HornikSW89}
K.~Hornik, M.~Stinchcombe, and H.~White, ``Multilayer feedforward networks are
  universal approximators,'' \emph{Neural networks}, vol.~2, no.~5, pp.
  359--366, 1989.

\bibitem{Infeld1945On}
G.~E. Uhlenbeck and L.~S. Ornstein, ``On the theory of the brownian motion,''
  \emph{Physical review}, vol.~36, no.~5, p. 823, 1930.

\bibitem{LillicrapHPHETS15}
T.~P. Lillicrap, J.~J. Hunt, A.~Pritzel, N.~Heess, T.~Erez, Y.~Tassa,
  D.~Silver, and D.~Wierstra, ``Continuous control with deep reinforcement
  learning,'' \emph{arXiv preprint arXiv:1509.02971}, 2015.

\bibitem{SilverLHDWR14}
D.~Silver, G.~Lever, N.~Heess, T.~Degris, D.~Wierstra, and M.~A. Riedmiller,
  ``Deterministic policy gradient algorithms,'' in \emph{Proceedings of the
  31th International Conference on Machine Learning, {ICML} 2014}, pp.
  387--395.

\bibitem{Barlacchi2015AMD}
G.~Barlacchi, M.~De~Nadai, R.~Larcher, A.~Casella, C.~Chitic, G.~Torrisi,
  F.~Antonelli, A.~Vespignani, A.~Pentland, and B.~Lepri, ``A multi-source
  dataset of urban life in the city of milan and the province of trentino,''
  \emph{Scientific data}, vol.~2, no.~1, pp. 1--15, 2015.

\bibitem{RN16}
\BIBentryALTinterwordspacing
wigle.[n.d.], ``Wireless basestation dataset.'' [Online]. Available:
  \url{https://wigle.net/.}
\BIBentrySTDinterwordspacing

\bibitem{BergstraB12}
J.~Bergstra and Y.~Bengio, ``Random search for hyper-parameter optimization,''
  \emph{J. Mach. Learn. Res.}, vol.~13, pp. 281--305, 2012.

\bibitem{kim2011dynamic}
H.-W. Kim, J.-H. Lee, Y.-H. Choi, Y.-U. Chung, and H.~Lee, ``Dynamic bandwidth
  provisioning using arima-based traffic forecasting or mobile wimax,''
  \emph{Computer Communications}, vol.~34, no.~1, pp. 99--106, 2011.

\bibitem{DeruyckJM14}
M.~Deruyck, W.~Joseph, and L.~Martens, ``Power consumption model for macrocell
  and microcell base stations,'' \emph{Transactions on Emerging
  Telecommunications Technologies}, vol.~25, no.~3, pp. 320--333, 2014.

\bibitem{LiuNX16}
C.~Liu, B.~Natarajan, and H.~Xia, ``Small cell base station sleep strategies
  for energy efficiency,'' \emph{IEEE Transactions on Vehicular Technology},
  vol.~65, no.~3, pp. 1652--1661, 2015.

\end{thebibliography}

\end{document}